\documentclass[12pt]{article}
\usepackage{fullpage}

\usepackage{etoc}

\usepackage{natbib}
\usepackage{amsmath}
\usepackage{amssymb}
\usepackage{amsthm}
\usepackage{mathtools}
\mathtoolsset{showonlyrefs}
\usepackage{thmtools,thm-restate}
\usepackage{zref-clever}
\zcsetup{cap=true,nameinlink=true}
\usepackage[pdfencoding=auto, psdextra]{hyperref}
\hypersetup{
	colorlinks,
	citecolor=blue,
	filecolor=black,
	linkcolor=black,
	urlcolor=black
}

\usepackage{graphicx}
\graphicspath{ {./img/} }
\usepackage{float}

\DeclareMathOperator{\iid}{\stackrel{iid}{\sim}}
\DeclareMathOperator{\st}{\text{ s.t. }}
\DeclareMathOperator{\period}{\text{.}}
\DeclareMathOperator{\comma}{\text{,}}
\DeclareMathOperator{\as}{\text{ as }}
\DeclareMathOperator{\since}{\text{since }}

\DeclareMathOperator{\where}{\text{ where }}
\DeclareMathOperator{\by}{\text{by }}
\DeclareMathOperator{\textif}{\text{if }}
\DeclareMathOperator{\otherwise}{\text{otherwise }}
\DeclareMathOperator{\textand}{\text{ and }}

\DeclareMathOperator{\for}{\text{ for }}

\newtheorem{theorem}{Theorem}
\newtheorem{lemma}{Lemma}
\newtheorem{remark}{Remark}
\newtheorem{corollary}{Corollary}
\newtheorem{proposition}{Proposition}
\newtheorem{definition}{Definition}

\newcommand*{\defeq}{=}
\let\rank\relax
\DeclareMathOperator{\rank}{rank}
\DeclareMathOperator{\CD}{CD}

\DeclareMathOperator{\OLS}{OLS}

\DeclareMathOperator{\SEM}{SEM}

\DeclareMathOperator{\supp}{supp}

\newcommand{\Rephat}{\hat{R}_{\ep}}

\newcommand{\Renv}{R_{\env}}
\newcommand{\Renvk}{R_{A_k}}

\newcommand{\Rdiff}{R_{\Delta}}

\newcommand{\betapa}{\beta_c}
\newcommand{\betapat}{\beta_{c}^{T}}

\newcommand{\betach}{\beta_d}

\newcommand{\B}{B}
\newcommand{\Bx}{B_X}

\newcommand{\betaols}{\beta_{\OLS}}
\newcommand{\betacd}{\beta_{\CD}}

\newcommand{\betahatlambda}{\hat{\beta}_\lambda}

\usepackage{bbm} % for indicator functions
\usepackage[bb=boondox]{mathalfa} % bb for numbers

\newcommand{\bzero}{\mathbb{0}}

\newcommand{\E}{\mathbb{E}} % expectation

\newcommand{\Rposzero}{[0,\infty)} % reals
\newcommand{\R}{\mathbb{R}} % reals
\newcommand{\Rp}{\mathbb{R}^p}
\newcommand{\Rpp}{\mathbb{R}^{p\times p}}

\newcommand{\Zlambda}{Z_{\lambda}}

\newcommand{\gGlambda}{G_{\lambda}^{\dagger}}

\newcommand{\Glambda}{G_{\lambda}}

\newcommand{\Gplus}{G_{+}}

\newcommand{\Gdelta}{G_{\Delta}}
\newcommand{\GDelta}{G_{\Delta}}

\newcommand{\Gdiff}{G_{\Delta}}

\newcommand{\Gsr}{G_{\Delta}^{1/2}}
\newcommand{\Gsrg}{G_{\Delta}^{g/2}}

\newcommand{\gbGlambda}{\hat{G}_{\lambda}^\dagger }

\newcommand{\bGlambda}{\hat{G}_{\lambda}}
\newcommand{\bZlambda}{\hat{Z}_{\lambda}}
\newcommand{\bG}{\hat{G}_{\Delta}}

\newcommand{\bGplus}{\hat{G}_{+}}

\newcommand{\Zdelta}{Z_{\Delta}}

\newcommand{\Zdiff}{Z_{\Delta}}

\newcommand{\Zplus}{Z_{+}}

\newcommand{\bZ}{\hat{Z}_{\Delta}}

\newcommand{\bZplus}{\hat{Z}_{+}}

\usepackage{relsize}

\DeclareMathOperator{\SG}{SG}

\DeclareMathOperator{\ep}{A}
\DeclareMathOperator{\normalep}{\tilde{A}}

\DeclareMathOperator{\N}{\mathcal{N}}
\DeclareMathOperator{\shifts}{\mathcal{A}}
\DeclareMathOperator{\inshifts}{\normalep\in\shifts}

\DeclareMathOperator{\envk}{A_k}
\DeclareMathOperator{\env}{A}
\DeclareMathOperator{\normalenv}{A_X}

\DeclareMathOperator{\obs}{0}
\DeclareMathOperator{\normalobs}{0}

\DeclareMathOperator{\insample}{\normalep\in\{\normalenv,\normalobs\}}

\DeclareMathOperator{\nepj}{n_A(S^A,j)}

\DeclareMathOperator{\Ae}{\normalenv}

\DeclareMathOperator{\EAeAe}{\E[\Ae\Ae^{T}]}

\DeclareMathOperator{\Xe}{X_{A}}

\DeclareMathOperator{\Xo}{X_{\obs}}

\DeclareMathOperator{\Ye}{Y_{A}}

\DeclareMathOperator{\Yo}{Y_{\obs}}

\DeclareMathOperator{\PX}{P_X}
\DeclareMathOperator{\PY}{P_Y}

\DeclareMathOperator{\bXep}{\mathbb{X}^{\ep}}

\DeclareMathOperator{\bXepSone}{\mathbb{X}^{\ep}_{S^A,1}}
\DeclareMathOperator{\bXepSzero}{\mathbb{X}^{\ep}_{S^A,0}}
\DeclareMathOperator{\bXepSj}{\mathbb{X}^{\ep}_{S^A,j}}

\DeclareMathOperator{\bYep}{\mathbb{Y}^{\ep}}

\DeclareMathOperator{\bYepSone}{\mathbb{Y}^{\ep}_{S^A,1}}

\DeclareMathOperator{\bYepSj}{\mathbb{Y}^{\ep}_{S^A,j}}

\DeclareMathOperator*{\argmin}{arg\,min}

\let\norm\relax
\newcommand{\norm}[1]{\lVert #1 \rVert}

\DeclareMathOperator*{\sign}{sign}

\usepackage{stackengine}

\DeclareMathOperator{\toP}{\overset{p}{\to}}

\DeclareMathOperator{\Pnep}{\mathbb{P}_{n_A}^{\ep}}

\DeclareMathOperator{\PSepone}{\mathbb{P}_{S^A,1}^{\ep}}

\DeclareMathOperator{\PSezero}{\mathbb{P}_{S^{\env},0}^{\env}}

\DeclareMathOperator{\PSozero}{\mathbb{P}_{S^{\obs},0}^{\obs}}

\DeclareMathOperator{\Rehatcv}{\hat{R}_{\env}^{S,1}}
\DeclareMathOperator{\Rephatcv}{\hat{R}_{\ep}^{S,1}}
\DeclareMathOperator{\Rohatcv}{\hat{R}_{\obs}^{S,1}}
\DeclareMathOperator{\Rdiffhatcv}{\hat{R}_{\Delta}^{S,1}}

\DeclareMathOperator{\betahatlambdacv}{\hat{\beta}_\lambda^{S,0}}

\usepackage{tikz}
\usetikzlibrary{quotes,shapes,decorations,arrows,calc,arrows.meta,fit,positioning}
\tikzset{
-Latex,auto,node distance =1 cm and 1 cm,semithick,
intervention/.style ={rectangle, draw=black},
random/.style ={circle, draw=black},
parameter/.style ={diamond, draw=black},
}

\begin{document}

\begin{center}

{\bf{\Large{Causal Regularization}\\{\normalsize A trade-off between in-sample and out-of-sample risk guarantees}}}

\vspace*{.2in}

{{
\begin{tabular}{cc}
Lucas Kania$^{\dagger}$  
& Ernst C. Wit$^{\diamond}$
\end{tabular}
}}

\vspace{.15in}

\begin{tabular}{c}
	$^\dagger$Department of Statistics 
    and Data Science, Carnegie Mellon University\\
	$^\diamond$Faculty of Informatics, Universit\`a della Svizzera italiana\\[0.12in]
\end{tabular}
\begin{tabular}{cc}
    \texttt{lucaskania@cmu.edu}, 
     \texttt{ernst.jan.camiel.wit@usi.ch} 
\end{tabular}
\phantom{\footnote{Email}}

\vspace{.15in}

\today

\end{center}

\begin{abstract}
Invariant prediction uses the prediction stability of causal relationships across different environments to identify causal variables. Conversely, using causal variables gives prediction guarantees even in out-of-sample data settings. In this paper, we investigate the identification of causal-like models from in-sample data that ensure out-of-sample risk guarantees when predicting a target variable from an arbitrary set of covariates. 

Ordinary least squares minimizes in-sample risk but offers limited out-of-sample guarantees, while causal models optimize out-of-sample guarantees at the expense of in-sample performance. We introduce a form of \textit{causal regularization} to balance these properties. In the population setting, higher regularization yields estimators with greater risk stability, albeit with increased in-sample risk. Empirically, however, there is a further trade-off to consider, as finite in-sample data reduced the ability to correctly identify models with high out-of-sample risk guarantees. We show how in such empirical settings the optimal causal regularizer can be found via cross-validation.  
\end{abstract}

\etocdepthtag.toc{mtchapter}
\etocsettagdepth{mtchapter}{section}
\etocsettagdepth{mtappendix}{none}
\etocsettagdepth{mtreferences}{section}
{
\renewcommand{\baselinestretch}{0}
\normalsize
\parskip=0em
\renewcommand{\contentsname}{\normalsize Table of contents}
\tableofcontents
}

\pagebreak

\section{Introduction}

In many modern inferential settings, the goal is to obtain models from in-sample data that are highly predictive for out-of-sample data. Heterogeneous in-sample data can hinder this goal by promoting overfitting \citep{hernanwhatif}. However, when this heterogeneity stems from distribution shifts that preserve the functional relationship between target and covariates, structural invariance implies that the causal model yields predictions robust to future shifts. This makes causal modelling valuable even for purely predictive tasks.

If the source of the distribution shift is unconfounded and known, instrumental variables \citep{dideleziv,imbensiv} can be constructed. Alternatively, if the sample can be split into sub-samples that isolate different instances of the distribution shift, the causal model provides invariant predictions regardless of the sub-sample. Under linearity and no confounding, \citet{icp} shows that regressing the target on its direct causes yields a model invariant to these distribution shifts. They proposed an algorithm that regresses the target and all possible subsets of covariates in each sub-sample and tests which regression returns the same model regardless of the sub-sample. Under various sources of heterogeneity, they prove that this approach identifies the causal model. Further extensions include \citet{buchholzLearningLinearCausal2023}, where the observed variables are allowed to result from a non-linear mapping applied to linearly related latent variables. However, the associated combinatorial search becomes infeasible in high dimensions.

\citet{irm} addresses the computational challenge by approximating the search, albeit with weaker identification guarantees under linearity \citep{riskofirm}. \citet{causaldantzig} introduce Causal Dantzig, which avoids the combinatorial search by modelling heterogeneity as an unobserved shift in a structural equation model (SEM). In this setting, the covariance between covariates and residuals under the causal model remains invariant across datasets. Given two datasets under sufficiently different shift distributions, the difference in these covariances forms a moment condition that identifies the causal model. If the solution is not unique, the authors propose an $L_1$ penalty to select a sparse approximation of the causal model. Thus, they obtain a biased estimator of the causal model, which, under some stringent assumptions on the regularization parameter, can provide guarantees regarding the distance between the estimator and the target. Recently \citet{polinelliGeneralisedCausalDantzig2024} extended the causal Dantzig to generalized linear models by exploiting the fact that for the causal model the expected risk based on the Pearson residuals remains invariant across distribution shifts.

In another development, \citet{rojas} show that the prediction invariance of the causal model implies that it minimizes the maximum risk over all possible distribution shifts. Following up on this,  \citet{anchorreg} show that the out-of-sample risk depends on the correlation between the residuals and the observed variable generating the additive distribution shift. The more uncorrelated they are, the stronger the out-of-sample risk guarantees under unseen distribution shifts, whereby the causal model provides the best guarantees. The authors propose a regularized estimator, called \emph{anchor regression}, based on the amount of correlation between the residuals and the shift variable. The out-of-sample guarantees and finite sample bounds of all regularized models follow from the regulated correlation. Ideally, a regularized model should be chosen based on subject matter knowledge about the maximum future shift of the data. If it is not known, cross-validation is recommended. Follow-up work extended the method to use noisy versions of the variable generating the distribution shift \citep{Oberst21} and discrete and censored outcomes \citep{Kook21}. Alternatively, if structural knowledge about the dependence structure among the variables is available, then methods advocated by  \cite{subbaswamy2018counterfactual} and \cite{subbaswamyUnifyingCausalFramework2022} allow precise characterization of which models are invariant to general interventions of the underlying data-generating process. The disadvantage of all the above methods is that they assume knowledge, e.g., about the shifts or dependence structure, that in most applied scenarios is unavailable. 

In this work, we consider the multiple datasets setting used in \citep{causaldantzig} without access to the random variable generating the distribution shift or knowledge of the dependencies between the observed random variables. We propose \emph{causal regularization} for obtaining estimators that progressively decrease the risk bound on shifted out-of-sample datasets, akin to anchor regression \citep{anchorreg}. Unlike it, the source of the distribution shift is unknown, and the proposed estimands are identifiable even under weak distribution shifts. We provide finite sample risk bounds for all regularized models and prove the adequacy of cross-validation and data splitting for attaining these bounds.

In an earlier version of this work, causal regularization evaluated the performance of parameters across datasets by comparing their risk across datasets. \citet{kennerbergWorstriskMinimizationGeneralized2024,kennerbergOptimalWorstriskMinimization2024,kennerbergFunctionalWorstRisk2025} and \citet{shenCausalityorientedRobustnessExploiting2025} extended this framework to accommodate any number of datasets and random structural equations. However, risk comparisons across datasets may yield inconsistent estimators when key assumptions break down. To address this limitation, we adopt a gradient-based risk comparison across datasets, which we explain in \zcref{sec:causal_regularization}. 

\section{Structural equation model with shifts}

%Studies with data from well-designed randomized interventions are able to extract causal parameters \citep{fisher1935design}. In this paper, we assume we have much more unstructured data.

In this section, we assume that the data generation process is given by a system of linear structural equations (SEM), where the shift across different distributions is given by an exogenous random variable. We introduce the key ideas from the causal Dantzig \citep{causaldantzig} and anchor regression \citep{anchorreg} and explain how they relate to causal regularization. Henceforth, if $A$ is a random variable, we use $\E[A]$ to denote the expectation of $A$.

\pagebreak
\begin{definition}[\bf Structural equation model]\label{def:sem} $V_A = \SEM(A)$ is defined as the stochastic solution of \begin{equation}\label{eq:semdef}
V_A= \begin{bmatrix}Y_A \\ X_A\end{bmatrix} \textand V_A  = \underbrace{B}_{\substack{\text{constant}\\\text{structure}}} \cdot\  V_A \ + \ \underbrace{\epsilon}_{\text{noise}} \ +  \underbrace{A}_{\substack{\text{distribution}\\\text{shift}}} \where B = \begin{bmatrix}
    0 & \betapat \\
    \betach & \Bx
\end{bmatrix}
\end{equation} and $V_A, \epsilon, A \in \mathbb{R}^{p+1}$ and $X_A\in\mathbb{R}^p$ are random vectors and $Y_A\in\mathbb{R}$ a random variable. The matrix $\Bx \in \Rpp$ consists of the interactions among the covariates $X_A$, the vector $\betapa \in \Rp$, called the causal parameters, describes the causal effects of $X_A$ on the target $Y_A$, and the vector $\betach \in \Rp$ are the downstream effects of $Y_A$ on $X_A$.

The noise $\epsilon$ has the same second moment irrespective of the shift, i.e., $\E[\epsilon\epsilon^{T}]=\Sigma$. Additionally, the random shift $A$ is assumed to be uncorrelated with the noise, i.e., $\E[A\epsilon^{T}]=0$ and to have a second moment $\E[AA^T]<\infty$. Finally, $I-\B$ is required to be non-singular in order for $V_A$ to be uniquely defined. \end{definition}

We emphasize that we impose no additional assumptions on $\Sigma$, which allows for correlation arising from unmeasured confounding. The non-singularity of $I-B$ holds for acyclic graphs \citep{causaldantzig}, and can also hold in a limiting sense for certain cyclic graphs; see Appendix 8.1 of \citet{anchorreg}. Finally, we note that when the distribution shift is generated by an exogenous intervention, the assumption that the noise remains uncorrelated with the shift holds. This is the case of all the experiments that we study in this work, detailed in \zcref{sec:lighttunnel,sec:additional_experiments}.

Henceforth, when clear from context, we use $A_X$ and $A_Y$ to refer to the entries of the random vector $A$ associated with the $p$-dimensional random variable $X_A$ and the univariate random variable $Y_A$, respectively. Let $\mathcal{A}$ be the set of all random vectors in $\mathbb{R}^{p+1}$ that have a finite second moment $\mathcal{A} = \left\{A\in\mathbb{R}^{p+1}: E[AA^{T}]<\infty\right\}$, and define the following second moments for any $[Y_A, X_A]^T=\SEM(A)$ for any $A \in \mathcal{A}$: \[G_A=\E[X_AX_A^{T}] \mbox{  and  } Z_A=\E[X_AY_A].\] 

\subsection{In-sample data setting}

We assume that we have data available an \emph{observational distribution}, i.e.  $[Y_0,X_0]^T=\SEM(0)$, and a \emph{shifted distribution}, i.e. $[Y_A,X_A]^T=\SEM(A) \st A\not\equiv0 \textand A \in \mathcal{A}$. Thus, we have access to the target and covariates but not the random variable $A$ that generates the distribution shift, contrary to what is assumed in \citet{anchorreg}. \zcref[S]{fig:graph} displays a prototypical graphical model that satisfies \zcref{def:sem}. 

\begin{figure}[ht]
  \begin{minipage}[c]{0.4\textwidth}
  \centering
  \begin{tikzpicture}
  \node[random] (X1)  {$X_1$};
  \node[random] (Y) [right = 1cm of X1]  {$Y$};
  \node[random] (X2)[right = 1cm of Y] {$X_2$};
  \node[random] (C) [above = 0.75cm of Y] {$U$};
  \node[random] (A) [below = 0.75cm of Y] {$A$};

  \path[draw,thick,->]
  (A) edge (X1)
  (A) edge (X2)
  (X1) edge["$\betapa$"] (Y)
  (Y) edge["$\betach$"]  (X2)
  (C) edge (X1)
  (C) edge (Y)
  (C) edge (X2);
  \path[draw,dashed,->]
  (A) edge["Id."] (Y);
  \end{tikzpicture}
  \end{minipage}\hfill
  \begin{minipage}[c]{0.6\textwidth}
    \caption{Pictorial representation of a shifted distribution. The target and covariates can be arbitrarily confounded. However, the unmeasured confounding random variable $U$ cannot be correlated with the source of the distribution shift $A$. In order to identify the causal parameters $\betapa$, the target cannot be shifted; see \zcref{prop:causal_identification}. That is, there should be no arrow from $A$ to $Y$. Nevertheless, this is not a requirement to obtain parameters with out-of-sample guarantees; see \zcref{lemma:decomposition}. The observational distribution corresponds to the same graph but removing the node $A$.}
    \label{fig:graph}
  \end{minipage}
\end{figure}

\subsection{Causal Dantzig}
The goal is to use observational and shifted distributions to learn a linear model that provides good performance on some unobserved distribution $\left(Y_{\tilde{A}}, X_{\tilde{A}}\right)=\SEM(\tilde{A})$ where $\tilde{A} \in \mathcal{A}$. In other words, from the observed distributions, we would like to obtain linear parameters that provide good predictions on data under a different distribution shift. We define the risk of linear parameters $\beta$ for the distribution $(Y_{\tilde{A}},X_{\tilde{A}})$ as \begin{equation}
R_{\tilde{A}}(\beta) = \E[(Y_{\tilde{A}}-\beta^TX_{\tilde{A}})^2] \period
\end{equation} 

If the observed $A$ is not a strong shift, meaning that both the shifted and observational distributions, $A$ and $0$ respectively, are fairly similar, it makes intuitive sense to pool the data and minimize the total risk. Here, we define an OLS estimand to be any minimizer of the pooled risk across distributions  \begin{equation}\label{eq:ols}
\beta_{\OLS} \in \argmin_{\beta \in \mathbb{R}^p} R_+(\beta) \where R_+(\beta)=R_A(\beta)+R_0(\beta)
\end{equation} There is a unique solution if and only if $G_+=G_A+G_0$ is non-singular. However, if the shifted and observational distributions are substantially different, OLS might make a large error in one of the two distributions. Potentially, this could imply that the overall in-distribution performance of OLS is not indicative of the performance on future data where the distribution shift is stronger.

If one assumes that the target variable is not shifted, i.e. $A_Y=0$ --- which is known as the \emph{exclusion restriction} in the potential outcomes framework \citep{potentialoutcomes} --- then the risk of the causal parameters $R_{\tilde{A}}(\betapa)$ remains invariant over $\tilde{A} \in \mathcal{A}$. In other words, the risk of the causal parameters is invariant to all shifts $A\in\mathcal{A}$. Other linear parameters do not have this property since for any $\beta\not=\betapa$, there exists a sequence of shifts such that the risk under $\beta$ is arbitrarily bad, i.e., $\exists \{\envk\}_{k=1}^{\infty} \in \mathcal{A} \st \lim_{k\to\infty} \Renvk(\beta) = \infty$. Hence, if the causal parameters could be estimated from in-sample data, the out-of-sample risk would be constant under them, which would be the best possible guarantee for arbitrary strong distribution shifts. Thus, we might use the risk invariance property to identify the causal parameters. 

Let a causal Dantzig estimand be any solution of the risk difference minimization \begin{equation}\label{eq:causaldantzig}
\beta_{\CD} \in \argmin_{\beta \in \mathbb{R}^p}  |R_\Delta(\beta)|  \where R_\Delta(\beta)=R_A(\beta)-R_0(\beta) .
\end{equation} 
The causal Dantzig estimand satisfies the following optimality equation,  as shown in \zcref{lemma:minimimum_cd_under_SEM} in \zcref{sec:causal_dantzig}, 
\begin{equation}\label{eq:causaldantzigmoment}
G_\Delta \betacd = Z_\Delta
\end{equation} where $G_\Delta=G_A-G_0$ captures the shift in the covariates across distributions, and $Z_\Delta=Z_A-Z_0$ captures the shift in the correlation between the target and the covariates. Without extra assumptions, it holds that \begin{equation}
\Gdelta \betacd = G_\Delta \betapa + M\left(\betach \E[A_Y^2] + \E[A_XA_Y]\right)
\end{equation} where $M=\left((I-\Bx)-\betach\betapat\right)^{-1}$. Thus, it can be noted that to identify the causal parameters via the causal Dantzig estimand, two sufficient conditions are the invertibility of $G_\Delta$ and the non-intervention of the target, i.e. $A_Y=0$. The following proposition summarizes that statement. The proof can be found in \zcref{sec:causal_dantzig}. 
\begin{proposition}[\bf Identification of causal parameters]\label{prop:causal_identification}
Assume an observational structural equation model $V_0=\SEM(0)$ and a shifted one $V_A=\SEM(A)$, in which the target is not shifted, i.e.,  $A_Y=0$, and the second moment matrix $E[A_XA_X^T]$ is full-rank. Then the causal Dantzig estimand is unique and coincides with the causal parameters \[\beta_{CD}=\betapa.\]
\end{proposition} %
%
% In summary, if we expect future distributions to be similar to the ones observed, then we would use an OLS estimand. Conversely, if future data suffers from stronger distribution shifts, the causal parameters are an interesting inferential target, and a causal Dantzig estimand is a sensible choice. If future data is, however, neither identically distributed nor completely dissimilar from the one observed, one can guess that interpolating between the OLS and causal Dantzig estimands might provide some parameters with better risk under mild distribution shifts. The next section explores such idea and shows how interpolating between these two estimands can be formally connected to out-of-sample risk guarantees.

In summary, when future distributions resemble those observed, the OLS estimand is appropriate. If stronger distribution shifts are expected, targeting the causal parameters becomes meaningful, and the causal Dantzig estimand offers a suitable alternative. In scenarios where future data is neither identically distributed nor entirely different, interpolating between the OLS and causal Dantzig estimands may yield parameters with improved risk under mild shifts. The next section develops this idea and establishes a formal connection between such interpolation and out-of-sample risk guarantees.

\section{Causal regularization}\label{sec:causal_regularization}

\citet{causaldantzig} proposed regularizing the causal Dantzig with an $\ell_1$ penalty when the causal parameters are not identifiable. Instead, we propose regularizing the causal Dantzig towards the OLS solution. It turns out that this can provide formal guarantees on the out-of-sample risk. % Furthermore, such guarantees do not require the identifiability of the causal parameters.

\begin{definition}[\bf Causal regularization]\label{def:causalregularization} Let $\lambda \in [0,\infty)$, a causal regularizer $\bar{\beta}_\lambda$ is defined as any solution to the following optimization objective \begin{equation}\label{eq:causalregularization}
\bar{\beta}_\lambda \in \argmin_{\beta \in \mathbb{R}^p} \frac{1}{2}R_+(\beta)+\frac{\lambda}{2}|R_{\Delta}(\beta)|
\end{equation}\end{definition} 
To understand what guarantees it provides, we define the set $C_\lambda$ of all distribution shifts that happen in the direction of the unobserved distribution shift $A$ with a certain $\lambda$ strength.

\begin{definition}[\bf Set of $\lambda$-times stronger shifts]\label{def:restrictedenvs} For a fixed shift-distribution $A \in \shifts$, let $\mathcal{C}_\lambda$ be the set of shifts that are $\lambda$-times stronger than $A$, i.e.,
\begin{align}
\mathcal{C}_\lambda = \left\{\inshifts ~\st~\begin{cases}
   \E[\tilde{A}\tilde{A}^T] \preceq \frac{1+\lambda}{2} \cdot \E[AA^T]  &\textif \lambda \in \Rposzero\\
   \supp(\tilde{A})=\supp(A)  &\textif \lambda = \infty
\end{cases}  \right\}
\end{align} where  $N \preceq M$  if $M - N$ is positive semi-definite and $\supp(A)=\{i \st A_i\not=0\}$. \end{definition}

As $\lambda$ increases, the set $C_\lambda$ of distribution shifts grows until it includes all distribution shifts that share the support with the unobserved distribution shift $A$. It holds that any causal regularizer $\bar{\beta}_\lambda$ controls the worst risk under any distribution shift on $\mathcal{C}_\lambda$. 

\begin{lemma}[\bf Worst risk decomposition]\label{lemma:worst_risk_decomposition}
Let $\lambda \geq 0$, then the worst risk in the out-of-sample set $\mathcal{C}_\lambda$ is precisely equal to a linear combination of the pooled risk and the risk difference in the two observed environments,
\begin{equation}
\sup_{\tilde{A} \in \mathcal{C}_\lambda} R_{\tilde{A}}(\bar{\beta}_\lambda) = \argmin_{\beta \in \mathbb{R}^p}\frac{1}{2}R_+(\beta) +\frac{\lambda}{2}|R_\Delta(\beta)|
\end{equation}
\end{lemma} 
The proof is deferred to \zcref{appx:decomposition}. Note that the above guarantee does not require the assumptions needed for identifying the causal parameters as in \zcref{prop:causal_identification}. Even if the target $Y$ is shifted $A_Y\neq 0$, and the distribution shift does not affect every covariate, then the causal regularizer $\bar{\beta}_\lambda$ still provides out-of-sample guarantees, even though the causal parameters cannot be identified.

The reason why regularizing towards the risk difference provides out-of-sample guarantees is that the risk difference measures the $L_2$ projection of the residuals to the linear span of the source of the distribution shift, as stated in the following proposition.
\begin{restatable}{proposition}{ProjectionA}
Let $[Y_A,X_A]^T=\SEM(A)$ and $[Y_0,X_0]^T=\SEM(0)$, then \begin{equation}\label{eq:residual_projection}
R_\Delta(\beta)=\E[(\E[Y_A-X_A^{T}\beta|A])^2]
\end{equation}
\end{restatable} Therefore, as $\lambda$ is increased in \zcref{def:causalregularization}, the risk difference $R_\Delta(\bar{\beta}_\lambda)$ decreases, and consequently, the residuals under $\bar{\beta}_\lambda$ become more uncorrelated with the source of the unobserved distribution shift $A$. That is, the predictions of the linear parameters $\bar{\beta}_\lambda$ are less correlated with the distribution shift and hence have better performance on unobserved distributions where the distribution shift $\tilde{A}$ happens in the same direction as $A$. We note that \citet{anchorreg} regularizes the OLS estimand using the right-hand side of equation \ref{eq:residual_projection} since they assume access to $A$. Thus, causal regularization can be seen as an attempt to perform anchor regression when the random variable producing the distribution shift is unobserved.

As noted, any solution to \eqref{eq:causalregularization} ensures out-of-sample risk guarantees. To maintain the practicality of the method and avoid introducing additional parameters, we choose the solution with the smallest $\ell_2$ norm: \begin{equation}\label{eq:consistency_issue}
\dot{\beta}_\lambda \in \argmin_{\beta \in M_\lambda}\norm{\beta}_2 \st M_\lambda=\argmin_{\beta\in\Rp} \frac{1}{2}R_+(\beta)+\frac{\lambda}{2}|R_\Delta(\beta)|.
\end{equation} If the risk difference does not vanish at the solution,  then the solution is unique and admits a closed-form expression: \begin{equation}
\dot{\beta}_\lambda = G_{\tilde{\lambda}}^\dagger Z_{\tilde{\lambda}} \where \tilde{\lambda}=\lambda \cdot \sign\left(R_\Delta(\dot{\beta}_\lambda)\right) \comma
\end{equation} 
where $G_\lambda = G_+ + \lambda \cdot G_\Delta$, $Z_\lambda = Z_+ + \lambda \cdot  Z_\Delta$ and $G_\lambda^\dagger$ is the Moore–Penrose pseudo-inverse of $G_\lambda$; see \zcref{thm:penrose} in the appendix. However, if the risk difference vanishes at the solution $R_\Delta(\dot{\beta}_\lambda) = 0$, the solution satisfies $\dot{\beta}_\lambda = G_{r}^\dagger Z_{r} \text{ for some } r \st |r|\leq \lambda$ but it is not guaranteed to be unique. The issue stems from the fact that the regularizer is not convex in general. Consequently, there might be two solutions. Thus, producing a consistent estimator when the risk difference is arbitrarily close to zero is challenging if the conditions of \zcref{def:sem} are not satisfied.

This consistency issue, rooted in non-identifiability, also affects estimators proposed in multi-dataset extensions of \eqref{eq:consistency_issue}. To mitigate this, \citet{kennerbergWorstriskMinimizationGeneralized2024,kennerbergOptimalWorstriskMinimization2024,kennerbergFunctionalWorstRisk2025} assume knowledge of which dataset is observational and rely on \zcref{def:sem} to establish identifiability. Alternatively, \citet{shenCausalityorientedRobustnessExploiting2025} avoid assuming \zcref{def:sem} but instead impose strong restrictions on the second moments of the observed data to ensure identifiability. 

To guarantee identifiability without relying on \zcref{def:sem} or imposing strong restrictions on the moments of the observed data, we consider the strategy followed by \citet{irm} and regularize the OLS towards the optimality condition of the causal Dantzig \eqref{eq:causaldantzigmoment}. That is, we use the gradient of the risk difference $\nabla_\beta R_\Delta(\beta)=2(\Gdelta\beta-\Zdelta)$  rather than the risk difference. \begin{definition}[\bf Causal regularization via optimality condition]\label{def:causalregularizationviaopt} Let $\lambda \in [0,\infty)$, a causal regularizer $\beta_\lambda$ is defined as any solution to the following optimization objective \begin{equation}\label{eq:causalregularizationviaopt}
\beta_\lambda = \argmin_{\beta \in M_\lambda}\norm{\beta}_2 \where M_\lambda=\argmin_{\beta\in\Rp} \ell_\lambda(\beta)
\end{equation}and $\ell_\lambda(\beta)=\frac{1}{2}R_+(\beta)+\frac{\lambda}{2}\norm{G_\Delta\beta-Z_\Delta}_2^2$. \end{definition} Since the regularizer is convex, the minimum norm solution is always guaranteed to be unique \citep{planitz1979}, and it has a simple closed-form solution: \begin{align}
\beta_\lambda = G_\lambda^\dagger Z_\lambda,
\end{align} where \begin{equation}\label{eq:Ggamma}
G_\lambda = \begin{dcases}
G_+ + \lambda \cdot G_\Delta^{T}G_\Delta &\textif \lambda \in [0,\infty)\\
G_\Delta&\textif \lambda = \infty
\end{dcases},\ Z_\lambda = \begin{dcases}
Z_+ + \lambda \cdot G_\Delta^{T}Z_\Delta &\textif \lambda \in [0,\infty)\\
Z_\Delta&\textif \lambda = \infty
\end{dcases}
\end{equation} and for the last case, we used the fact that $
\beta_\infty = (G_\Delta^{T}G_\Delta)^\dagger G_\Delta^{T}Z_\Delta = G_\Delta^\dagger Z_\Delta
$, see \zcref{thm:penrose} in \zcref{appx:causal_regularization}. Note that $\beta_\lambda$ still interpolates the minimum norm OLS solution $\beta_0=\betaols=G_+^\dagger Z_+$ and the causal Dantzig $\beta_\infty=\betacd$. Finally, it provides an out-of-sample risk guarantee under \zcref{def:sem}. 
\begin{lemma}[\bf Control worst risk via optimality condition]\label{lemma:control_via_grad}
Let $[Y_A,X_A]^T=\SEM(A)$ and $[Y_0,X_0]^T=\SEM(0)$, there exists constants $K \geq 0$ and $r > 0$ such that \begin{equation}
\min_{\beta\in\Rp}\sup_{\tilde{A}\in\mathcal{C}_{\lambda/r}} R_{\tilde{A}}(\beta) \leq \frac{K}{r} \cdot \lambda + \min_{\beta\in\Rp} \ell_\lambda(\beta)
\end{equation} where $
K = \frac{1}{2}\cdot\min\{|R_{\Delta}(\beta)| : \beta \in \mathbb{R}^{p} \textand \Gdelta\beta=\Zdelta\}$. \end{lemma} The proof can be found in \zcref{appx:decomposition}.

\section{The plug-in estimator}

In this section, we study the finite sample guarantees and consistency of the plug-in estimators without assuming that that the data are defined by an underlying SEM. Henceforth, let $(\mathbb{X}_0,\mathbb{Y}_0) \in \R^{n_0\times(p+1)}$ be $n_0$ identically distributed observations and $(\mathbb{X}_A,\mathbb{Y}_A) \in \R^{n_A\times(p+1)}$ be $n_A$ identically distributed observations. Let $\hat{G}_A=\mathbb{X}_A^T\mathbb{Y}_A/n_A$ and $\hat{Z}_A=\mathbb{X}_A^T\mathbb{Y}_A/n_A$ be the plug-in estimators for $G_A$ and $Z_A$. Furthermore, define the plug-in estimators  $\hat{Z}_+ = \hat{Z}_A+\hat{Z}_0$ , $\hat{G}_+ = \hat{G}_A + \hat{G}_0$, $\hat{G}_\Delta = \hat{G}_A - \hat{G}_0$, and $\hat{Z}_\Delta = \hat{Z}_A-\hat{Z}_0$. Then, the plugin estimator of \eqref{eq:causalregularizationviaopt} is \begin{equation}\label{eq:causalRegularizationViaOptHat}
\hat{\beta}_\lambda = \hat{G}_{\lambda}^\dagger\hat{Z}_\lambda
\end{equation} where $\hat{G}_\lambda =  \hat{G}_+ + \lambda \cdot \hat{G}_\Delta^{T}\hat{G}_\Delta$ and $\hat{Z}_\lambda = \hat{Z}_+ + \lambda \cdot \hat{G}_\Delta^{T}\hat{Z}_\Delta$. If computing the pseudo-inverse in \eqref{eq:causalRegularizationViaOptHat} is impractical due to dimensionality, a simple gradient descent algorithm can approximate $\tilde{\beta}_{\lambda}$, see Proposition 1 of \citet{hastieSurprisesHighdimensionalRidgeless2022}. Insofar as second moments exist and $G_\lambda$ is positive definite, the plug-in estimator is consistent.

\begin{restatable}[\bf Consistency]{proposition}{consistency}\label{prop:consistency} Fix the dimension $p \in \mathbb{N}$. If $Z_\lambda < \infty$ and $G_\lambda$ is positive definite, then $\hat{\beta}_\lambda$ \eqref{eq:causalRegularizationViaOptHat} is a consistent estimator of $\beta_\lambda$ \eqref{eq:causalregularizationviaopt}.
\end{restatable}

The proof can be found in \zcref{appx:consistency}. The positive definite assumption is satisfied, for example, when $G_+$ is full-rank and $\lambda=0$, $G_\Delta$ is full-rank and $\lambda=\infty$, and whenever either one is full-rank and $\lambda \in (0,\infty)$. Consistency can be strengthened to concentration if the tails of $(Y_A,X_A)$ and $(Y_0,X_0)$ decay fast enough. In the following, we introduce sub-Gaussianity and prove the concentration of the plug-in estimator.

\begin{definition} $V \in \R$ is a sub-Gaussian with mean zero and proxy variance $\sigma^2$ if \begin{equation}
E\left[  e^{\gamma V} \right]  \leq e^{\gamma\cdot \sigma^2 / 2}\quad \forall \gamma \in \R,
\end{equation} which we denote by $V \in \SG(\sigma^2)$. Furthermore, $V \in \Rp$ is sub-Gaussian with mean zero and variance $\sigma$, if every one-dimensional projection is sub-Gaussian: \begin{equation}
u^{T}V \in \SG(\sigma^2)\quad \forall u \in \{u \in \Rp: \norm{u}_2=1\}.
\end{equation}
\end{definition} Most importantly, it includes bounded and Gaussian random variables. If $E[V]=0$ and $\norm{V}_\infty \leq L$ almost surely, then $V\in \SG(2L)$. Furthermore, if $V \sim \mathcal{N}(0,\Sigma)$, then $V \in \SG\left(\gamma_{\max}(\Sigma)\right)$ where $\gamma_{\max}(\Sigma)$ is the largest eigenvalue of $\Sigma$.

Finally, for $\hat{\beta}_\lambda$ to concentrate around $\beta_\lambda$, it is necessary for the spectrum of $\hat{G}^\dagger_\lambda$ to be close to that of $G^\dagger_\lambda$, which in turn requires $G_\lambda$ to be positive definite and the sample size to grow linearly with the dimension.

\begin{restatable}[\bf Concentration]{proposition}{Concentration}\label{prop:concentration}
Let $V_A=[Y_A,X_A]^T\in \SG(\sigma^2)$, $V_0=[Y_0,X_0]^T \in \SG(\sigma^2)$ and $n=\min(n_A,n_0)$. If $\gamma_{\min}(G_\lambda)>0$, there exists a positive constant $C$ such that whenever \begin{equation}
n \geq C \cdot \left[\max\left(\frac{\sigma}{\gamma_{\min}(G_\lambda)},\sigma,1\right)\right]^2 \cdot (p+\log(1/\delta)),
\end{equation} it holds that \begin{equation}
\ell_\lambda(\hat{\beta}_\lambda) \leq \inf_{\beta \in \mathbb{R}^p}\ell_\lambda(\beta) + C' \cdot \left[\max\left(\frac{1}{\gamma_{\min}(G_\lambda)},1\right)\right]^2 \cdot\sigma \cdot \sqrt{\frac{p+\log(1/\delta)}{n}}
\end{equation} with probability at least $1-\delta$, where $C'$ is a positive constant that depends on $\lambda$, $\norm{G_\lambda}$, $\norm{Z_\lambda}_2$, $\norm{G_\Delta}$ and $\norm{Z_\Delta}_2$.
\end{restatable}

The above lemma holds without any assumptions on the correlation structure between the random variables. Additionally, if \zcref{def:sem} holds, the worst out-of-sample risk is controlled by combining the \zcref{prop:concentration} with \zcref{lemma:control_via_grad}. The proof can be found in \zcref{appx:concentration}.

Note that if one can ensure that $\norm{\hat{\beta}_\lambda}_1$ remains bounded, the dependence on the dimension improves from $p$ to $\log p$. A straightforward way to achieve this is to impose a fixed $\ell_1$ constraint on the norm instead of relying on the minimum norm solution, whose norm may increase with the dimension. However, this approach requires tuning two parameters instead of one. We discuss this method in \zcref{appx:l1_penalty}, but for practical purposes, we continue to use the minimum norm solution due to its simplicity.

\section{Model selection}\label{sec:modelselection}

We have so far evaluated out-of-sample risk for a fixed regularization parameter $\lambda$. In this section, we show how to use in-sample data to select $\lambda$ with strong out-of-sample guarantees.

\begin{figure}[ht]
\centering
\includegraphics[width=\linewidth]{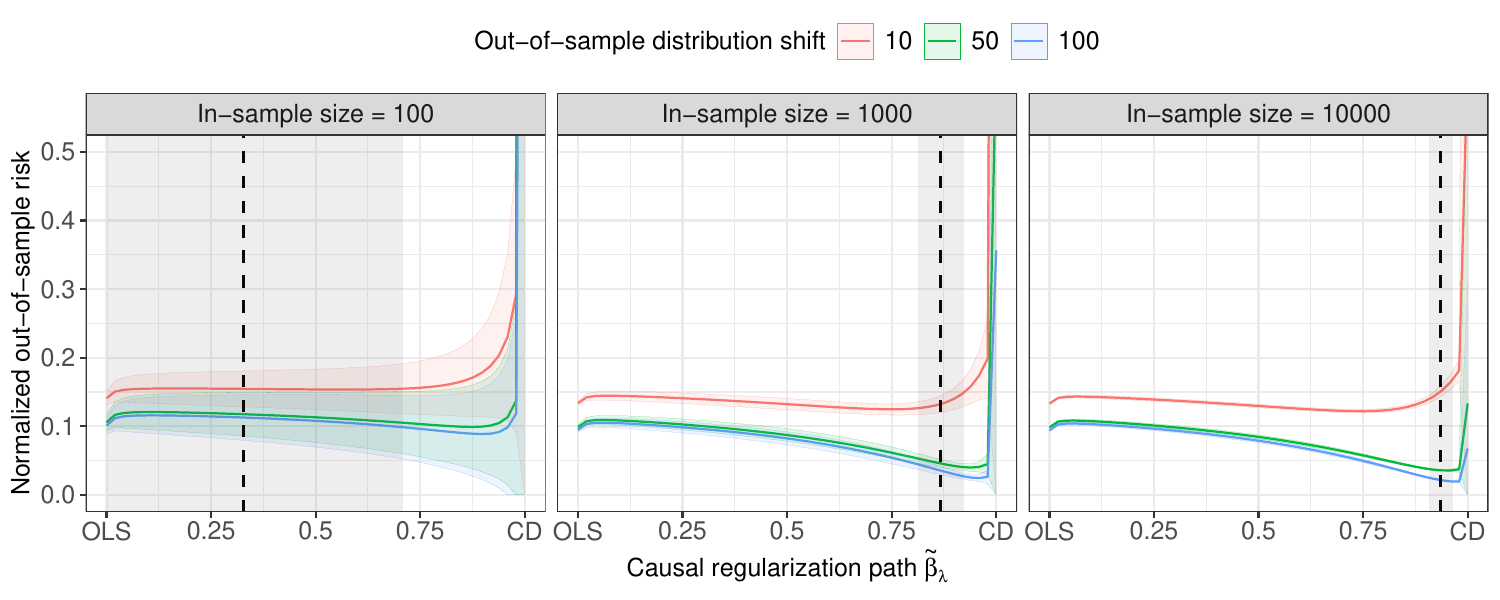}
\caption{Out-of-sample risk along the causal regularization path \eqref{eq:causalRegularizationViaOptHat} as sample size and distribution shift increase. The dashed vertical line marks the parameter chosen by cross-validation. Shaded areas represent one standard deviation across $1000$ trials. The causal regularization path has been normalized, see \eqref{eq:normalization}.}
\label{fig:increasing_sample_out_of_sample_risk}
\end{figure}

Consider data that satisfies \zcref{def:sem}, \zcref{fig:increasing_sample_out_of_sample_risk} illustrates the out-of-sample risk along the causal regularization path from OLS to the causal Dantzig. When the sample size is sufficiently large, a value of $\lambda$ exists that outperforms both OLS and the causal Dantzig under strong distribution shifts between in-sample and out-of-sample data. We approximate this phenomenon using in-sample data via \zcref{lemma:worst_risk_decomposition}, which states that under strong out-of-sample shifts, out-of-sample risk scales with the in-sample risk difference,
\begin{equation}\label{eq:in_out_sample_connection}
\forall \beta\in\Rp \quad \lim_{\lambda\to\infty}\sup_{\tilde{A} \in \mathcal{C}_\lambda} \frac{R_{\tilde{A}}(\beta)}{\lambda} = \frac{1}{2}|R_\Delta(\beta)|\period
\end{equation}
\zcref[S]{fig:increasing_sample_in_sample_risk} shows that the in-sample risk distribution serves as a reliable indicator of out-of-sample performance, see \zcref{fig:increasing_sample_out_of_sample_risk}, when sufficient observations and in-sample distribution shifts are present. Consequently, a simple approach is in order to select a regularization parameter is to split the data, fit causal regularization on the first half, and select the model minimizing the risk difference on the second half as a surrogate for out-of-sample risk.

\begin{figure}[tb]
\centering
\includegraphics[width=\linewidth]{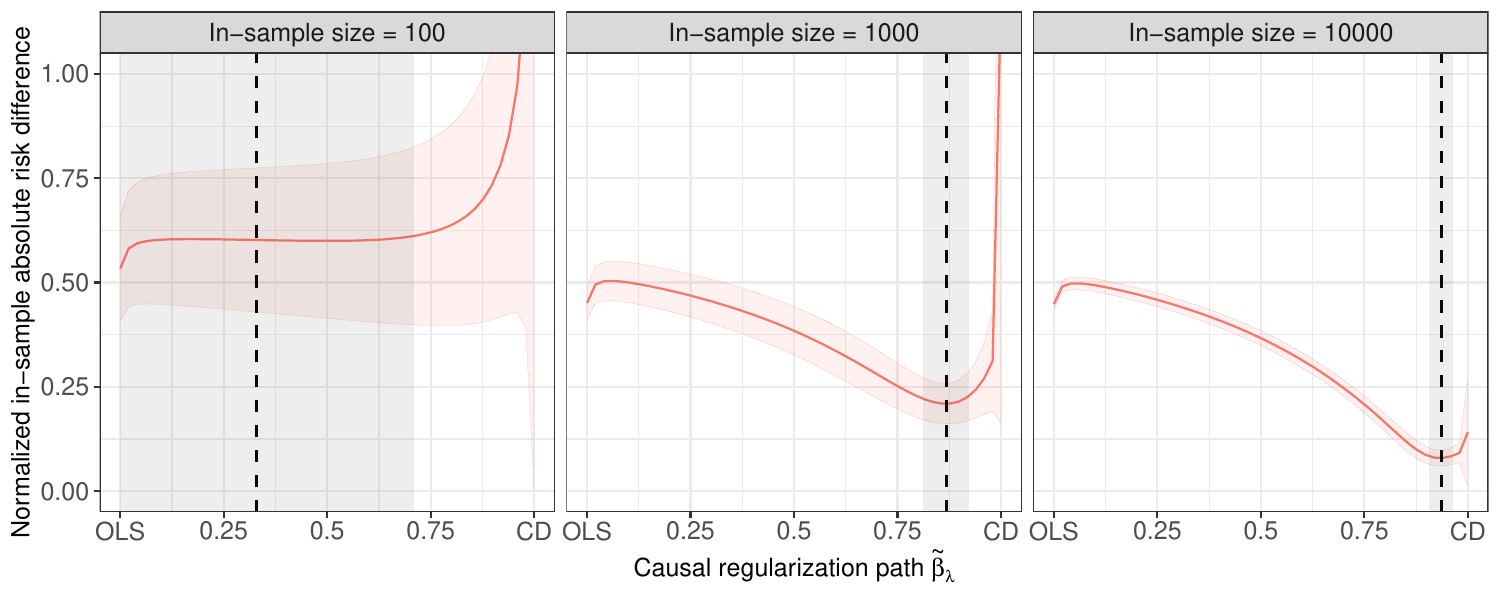}
\caption{Normalized in-sample absolute risk difference for the causal regularization path \eqref{eq:causalRegularizationViaOptHat} as the sample size increases. The dashed vertical line marks the parameter chosen by cross-validation. Shaded areas represent one standard deviation across $1000$ trials. The causal regularization path has been normalized, see \eqref{eq:normalization}.}
\label{fig:increasing_sample_in_sample_risk}
\end{figure}

Let $(\mathbb{X}^S_0,\mathbb{Y}^S_0) \in \R^{n_0\times(p 1)}$ and $(\mathbb{X}^S_A,\mathbb{Y}^S_A) \in \R^{n_A\times(p 1)}$ for $S\in \{0,1\}$ represent the two data splits. Define the empirical risk of $\beta$ on $(\mathbb{X}^S_A,\mathbb{Y}^S_A)$ as $\hat{R}^S_A(\beta)=\norm{\mathbb{Y}^S_A-\mathbb{X}^S_A\beta}_2^2 / n_A$. Let $\hat{\beta}_\lambda = \hat{\beta}_\lambda\left(\mathbb{X}^0,\mathbb{Y}^0\right)$ be the causal regularization fit at $\lambda$ using the first half of the data $S=0$. We choose $\lambda$ as
\begin{equation}\label{eq:model_selection}
\hat{\lambda} = \inf_{\lambda \in \Lambda} |\hat{R}_\Delta(\hat{\beta}_\lambda)|,
\end{equation}
where $\Lambda$ is a finite set and $\hat{R}_{\Delta}(\beta) = |\hat{R}^1_{A}(\beta)-\hat{R}^1_{0}(\beta)|$ is the risk difference on the second half. \zcref[S]{fig:increasing_sample_in_sample_risk} shows that this surrogate loss supports effective model selection.  When the sample size is too small to detect an in-sample distribution shift, see the first panel of \zcref{fig:increasing_sample_in_sample_risk}, the variance of the causal Dantzig is high, resulting in a high in-sample risk difference. Consequently, $\hat{\lambda}$ tends toward zero, corresponding to the OLS solution. In contrast, with sufficient data to detect a distribution shift, see the third panel of \zcref{fig:increasing_sample_in_sample_risk}, the causal Dantzig exhibits lower variance and achieves a smaller in-sample risk difference, so $\hat{\lambda}$ moves closer to one, selecting the causal Dantzig solution.

In \zcref{appx:model_selection}, we justify that either sample splitting or cross-validation will asymptotically behave like an oracle that has access to an infinite amount of data. The results follow from the work of \citet{cvvanderlaan}. We note that in all experiments, we use cross-validation rather than sample splitting to improve data efficiency. We also normalize $\lambda$ to the $[0,1]$ interval, which simplifies comparisons across experiments. Let
\begin{align}\label{eq:normalization}
\tilde{\beta}_{\lambda}=\tilde{G}_\lambda^\dagger \tilde{Z}_\lambda\quad \text{for } 0 \leq \lambda \leq 1,
\end{align}
where $\tilde{G}_\lambda = (1-\lambda)\cdot \hat{G}_+ + \lambda \cdot\hat{G}_\Delta^{T}\hat{G}_\Delta$ and $\tilde{Z}_\lambda = (1-\lambda)\cdot \hat{Z}_+ + \lambda\cdot  \hat{G}_\Delta^{T}\hat{Z}_\Delta$. Thus $\tilde{\beta}_0$ corresponds to the OLS and $\tilde{\beta}_1$ corresponds to the causal Dantzig estimator.

\section{Simulation studies}\label{sec:simulation}

We compare causal regularization to the causal Dantzig estimator and OLS. In all cases, we use cross-validation to select the regularization parameter for causal regularization.

\begin{figure}[!ht]
\centering

\begin{minipage}{0.3\textwidth}
\centering
\begin{tikzpicture}
\node[random] (x2)  {$X_2$};
\node[random] (x3) [right = 1cm of x2]  {$X_3$};
\node[random] (x1) [above left = 0.75cm of x3] {$X_1$};
\node[random] (y) [below left  = 0.75cm of x3] {$Y$};
\node[random] (u) [left  = 1cm of y] {$U$};
\node[random] (x4) [below = 1.4cm of x2] {$X_4$};
\node[random] (x5) [right  = 1cm of x4] {$X_5$};
\node[random] (x6) [below  = 0.75cm of x5] {$X_6$};

\path[draw,thick,->]
(u) edge[bend left=40] (x1)
(u) edge (x2)
(u) edge (y)
(u) edge (x4)
(u) edge[bend right=40] (x6)
(u) edge[out=105, in=75, looseness=2] (x3)
(u) edge[bend right=80] (x5)
(x1) edge (x2)
(x1) edge  (x3)
(x2) edge (x3)
(x2) edge (y)
(x3) edge (y)
(y) edge (x4)
(y) edge (x5)
%(x4) edge (x5)
(x5) edge (x6);
\end{tikzpicture}

\end{minipage}%%%
\begin{minipage}{0.7\textwidth}
\centering
% \scriptsize
\begin{align}
    \B=\begin{bmatrix}
        0 & \betapat \\
        \betach & \Bx
    \end{bmatrix} &\where \betapa = \begin{bmatrix}
        0\\1\\1\\0\\0\\0\\
    \end{bmatrix},\ \betach = \begin{bmatrix}
        0\\0\\0\\1\\1\\0\\
    \end{bmatrix} \textand 
\end{align}
\[ \Bx =  \begin{bmatrix}
        0&0&0&0&0&0\\
        1&0&0&0&0&0\\
        1&1&0&0&0&0\\
        0&0&0&0&0&0\\
        0&0&0&0&0&0\\
        0&0&0&0&1&0\\
    \end{bmatrix}\]

\end{minipage}

\caption{$\SEM(A)$ used for the simulation study in \zcref{sec:increasing_sample}. $U$ denotes an unmeasured confounder that affects all measured variables.}
\label{fig:semsimulations}
\end{figure}

\subsection{Increasing sample size and fixed dimension}\label{sec:increasing_sample}

We study causal regularization as the sample size increases and the out-of-sample distribution shift varies. This section focuses on a setting where the causal parameters are identifiable, allowing the causal Dantzig to outperform OLS for a large enough sample size. We defer the case with non-identifiable parameters to \zcref{appx:increasing_sample}.

We generate data from the structural equation model defined in \zcref{def:sem}, where the structural matrix $B$ has dimension $p=7$ as shown in \zcref{fig:semsimulations}. The unmeasured confounder, noise, and shift distributions follow
\begin{equation}\label{eq:dgp}
\epsilon \sim \N\left(0, I_p \right), U \sim \N\left(0,1\right), A_X \sim  \sqrt{\gamma_1}\cdot \N(0_p,I_p), \textand A_Y=0,
\end{equation} where $0_p$ is a $p$-dimensional column vector of zeroes and $I_p$ is the $p$-dimensional identity matrix. The target variable is not directly shifted, ensuring that the causal parameters are identifiable.

We create in-sample datasets with $n \in \{10^2,10^3,10^4\}$ and fix $\gamma_1=5$ in \eqref{eq:dgp}. We select the causal regularization parameter using 10-fold cross-validation and evaluate the risk of $\hat{\beta}_\lambda$ for $\lambda \in [0,1]$ on out-of-sample data drawn from \eqref{eq:dgp} with $\gamma_1 \in \{10,50,100\}$, representing distribution shifts 2, 10, and 20 times stronger than those observed in-sample. Each configuration is repeated over 1000 trials.

\zcref[S]{fig:increasing_sample_out_of_sample_risk} displays the out-of-sample risk across the causal regularization path as the sample size increases. On average, cross-validation selects a $\lambda$ close to the optimal. \zcref[S]{fig:increasing_sample_comparison} further supports this result by comparing the out-of-sample risk of OLS, the causal Dantzig, and the selected causal regularizer. For small samples, where in-sample distribution shifts are hard to detect, the selected regularizer performs similarly to OLS. As the sample size grows and the shift becomes more apparent, it approaches the causal Dantzig’s performance while maintaining lower variance by shrinking toward the OLS solution.

\begin{figure}[ht!]
\centering
\includegraphics[width=\linewidth]{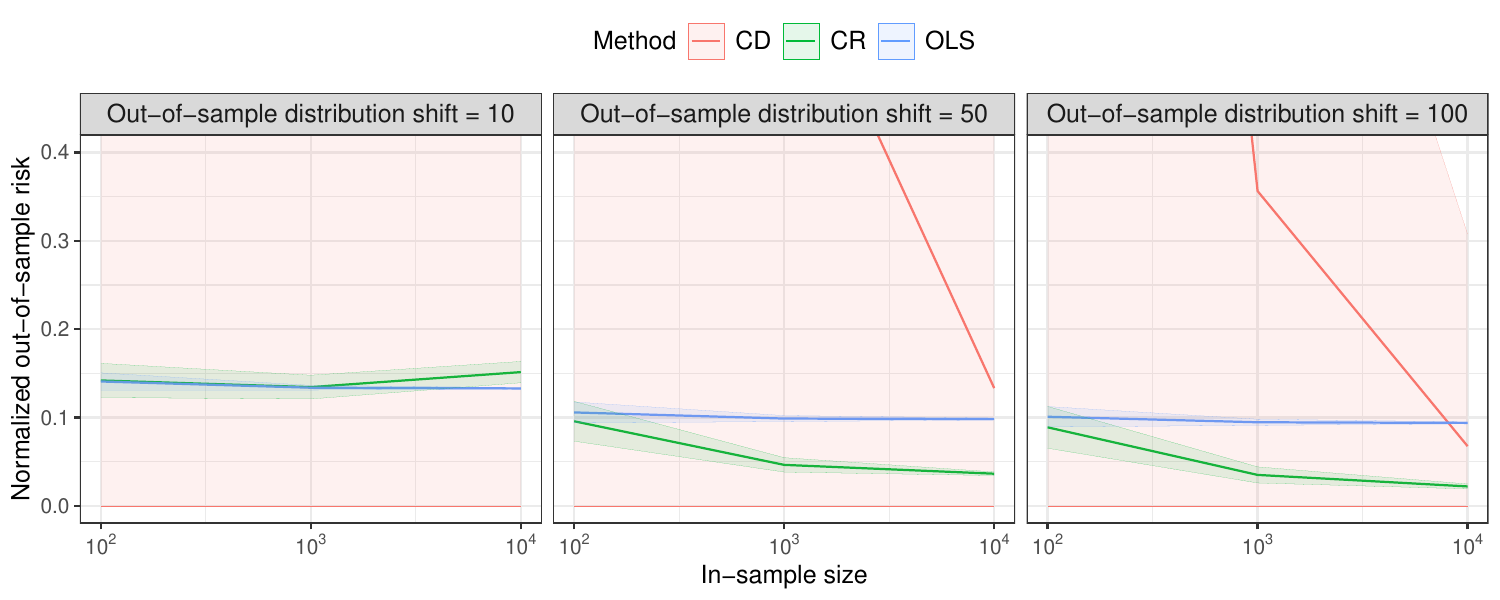}
\caption{Out-of-sample risk for ordinary least-squares, causal dantzig and causal regularization \eqref{eq:causalRegularizationViaOptHat} as the sample size and out-of-sample distribution shift increases. The target variable has not been directly shifted, i.e., $\gamma_2=0$ in \eqref{eq:dgp}. Solid lines indicate average performance, while shaded regions represent one standard deviation across $1000$ trials.}
\label{fig:increasing_sample_comparison}
\end{figure}

\subsection{Increasing dimension}\label{sec:increasing_dimension}

We study causal regularization in a high-dimensional setting. As in the previous section, we defer the case with non-identifiable causal parameters to \zcref{appx:increasing_sample}. Data is sampled from \zcref{def:sem}, with noise and distribution shifts defined by \eqref{eq:dgp}. The structural matrix $B$ is given by: \begin{equation}
B_{i,j}=\begin{cases}
\text{Bernoulli}(1/2) &\textif i\not=j\\
0 &\textif i=j
\end{cases} \for 1\leq i,j\leq p+1.
\end{equation} This setup allows cycles but excludes self-loops. The target is chosen to be the variable with the largest number of parents. Furthermore, as in \zcref{sec:increasing_sample}, we add a unmeasured confounding variable that follows a standard normal distribution and affects all measured variables. We generate in-sample datasets of varying sizes with $\gamma_1 = 5$ in \eqref{eq:dgp}. We tune the causal regularization using 10-fold cross-validation, and evaluate the risk of $\hat{\beta}_\lambda$ for $\lambda \in [0,1]$ on out-of-sample data generated with $\gamma_1 = 100$. We conduct 30 trials for each of 30 random graphs per dimension and sample size. 

\zcref[S]{fig:increasing_dim_comparison} presents the results as the dimension increases. Note that if the sample size is low or the dimension is high, causal regularization matches the performance of OLS since the in-sample distribution shift is not detectable. However, in low-dimensional settings with large sample sizes, causal regularization can improve on average over OLS.

\begin{figure}[ht!]
\centering
\includegraphics[width=\linewidth]{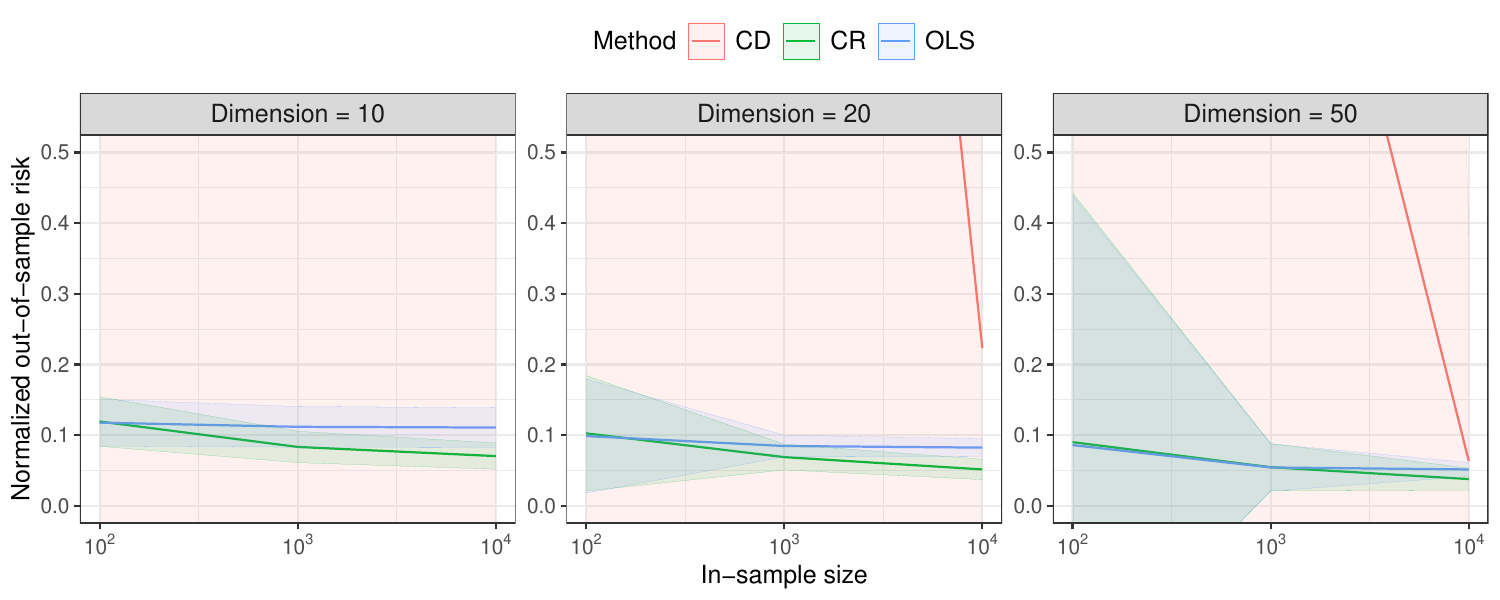}
\caption{Normalized out-of-sample risk for ordinary least-squares, causal Dantzig and causal regularization \eqref{eq:causalRegularizationViaOptHat} for varying dimension. Solid lines indicate average performance, while shaded regions represent one standard deviation across $30$ random graphs and $30$ trials per graph. In the third panel, the shaded regions for OLS and causal regularization overlap.}
\label{fig:increasing_dim_comparison}
\end{figure}

\section{Prediction under interventions in a light tunnel}\label{sec:lighttunnel}

We apply OLS, causal Dantzig, and causal regularization in a controlled experimental setting to assess whether the patterns observed in \zcref{sec:increasing_sample,sec:increasing_dimension} also emerge in empirical scenarios. In every experiment, we follow a holdout strategy by splitting the data into training and test sets. Using the training set, we select the causal regularization parameter $\lambda$ via cross-validation. We then compute the regularization path fitted on the training data and evaluate its risk on the test set. This allows us to compare the performance of OLS, causal Dantzig, and the tuned causal regularizer. We expect the selected causal regularizer to perform similarly to OLS when the training data does not exhibit a significant distribution shift. However, it should outperform OLS when there is a detectable in-sample distribution shift.

We use data from a light tunnel experiment introduced by \citet{gamellaCausalChambersRealworld2025}. The tunnel consists of a chamber with a controllable light source, two rotating linear polarizers, and sensors that record various physical quantities. The control inputs include the brightness of the red, green, and blue LEDs ($R$, $G$, $B$) and the angles of the polarizer frames ($\theta_1$, $\theta_2$). Outputs include infrared light intensity readings ($\tilde{I}_1$, $\tilde{I}_2$, $\tilde{I}_3$), visible light intensity readings ($\tilde{V}_1$, $\tilde{V}_2$, $\tilde{V}_3$), the electrical current drawn by the light source ($\tilde{C}$), and other measurements. \zcref{fig:chamber_graph} shows the causal graph for the experiment. Further details are provided in \citet{gamellaCausalChambersRealworld2025}.

\begin{figure}[ht!]
\centering
\begin{tikzpicture}[>=Stealth, node distance=1.6cm, every node/.style={align=center}]
% Top nodes
\node (RGB) at (0,0) {$(R,G,B)$};
\node (Ctilde) [right=1.5cm of RGB] {$\tilde{C}$};

% Middle row: I and V pairs
\node (I1) [below left=1.2cm and 3cm of RGB] {$\tilde{I}_1$};
\node (V1) [right=1.8cm of I1] {$\tilde{V}_1$};

\node (I2) [below=1.2cm of RGB] {$\tilde{I}_2$};
\node (V2) [right=1.8cm of I2] {$\tilde{V}_2$};

\node (I3) [below right=1.2cm and 3cm of RGB] {$\tilde{I}_3$};
\node (V3) [right=1.8cm of I3] {$\tilde{V}_3$};
\node (theta) [above=1.2cm of V3] {$(\theta_1, \theta_2)$};

% Bottom row: L groups
\node (L1) [below=1cm of V1] {$(L_{11}, L_{12})$};
\node (L2) [below=1cm of V2] {$(L_{21}, L_{22})$};
\node (L3) [below=1cm of V3] {$(L_{31}, L_{32})$};

% Arrows from RGB
\draw[->] (RGB) -- (Ctilde);
\draw[->] (RGB) -- (I1);
\draw[->] (RGB) -- (I2);
\draw[->] (RGB) -- (I3);
\draw[->] (RGB) -- (V1);
\draw[->] (RGB) -- (V2);
\draw[->] (RGB) -- (V3);

% theta -> I3 and V3
\draw[->] (theta) -- (I3);
\draw[->] (theta) -- (V3);

% L -> I and V
\draw[->] (L1) -- (I1);
\draw[->] (L1) -- (V1);

\draw[->] (L2) -- (I2);
\draw[->] (L2) -- (V2);

\draw[->] (L3) -- (I3);
\draw[->] (L3) -- (V3);

\end{tikzpicture}
\caption{Graphical representation of the structural equations underlying the light tunnel experiment.}
\label{fig:chamber_graph}
\end{figure}

We evaluate the performance of causal regularization in predicting the second light intensity measurement ($\tilde{I}_2$) under three settings: using only its causal parents ($R$, $G$, $B$); adding a few additional correlated variables ($\tilde{I}_1$, $\tilde{I}_3$, $\tilde{C}$, $\tilde{V}_1$, $\tilde{V}_2$, $\tilde{V}_3$); and including all variables from \zcref{fig:chamber_graph}.

\begin{figure}[ht!]
\centering
\includegraphics[width=\linewidth]{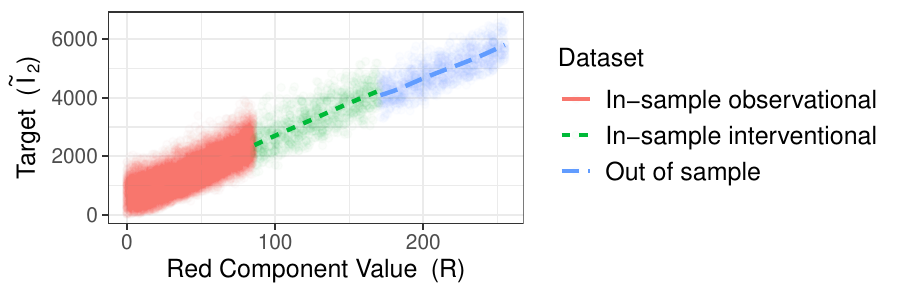}
\caption{Relationship between the target and the value of the red component across the in-sample and out-of-sample datasets.}
\label{fig:chamber_datasets}
\end{figure}

As training data, we use an observational dataset with $10000$ samples collected without interventions, together with data where the light sources were moderately intervened ($R, G, B \iid \text{Unif}(\{ 86, \dots, 170\})$). For testing, we use the dataset where the light sources underwent strong interventions ($R, G, B \iid \text{Unif}(\{ 171, \dots, 255\})$). Each interventional dataset contains $3000$ samples. \zcref[S]{fig:chamber_datasets} shows the linear relationship between the target and the red component; similar patterns appear for the blue and green components.

\begin{figure}[ht!]
\centering
\includegraphics[width=\linewidth]{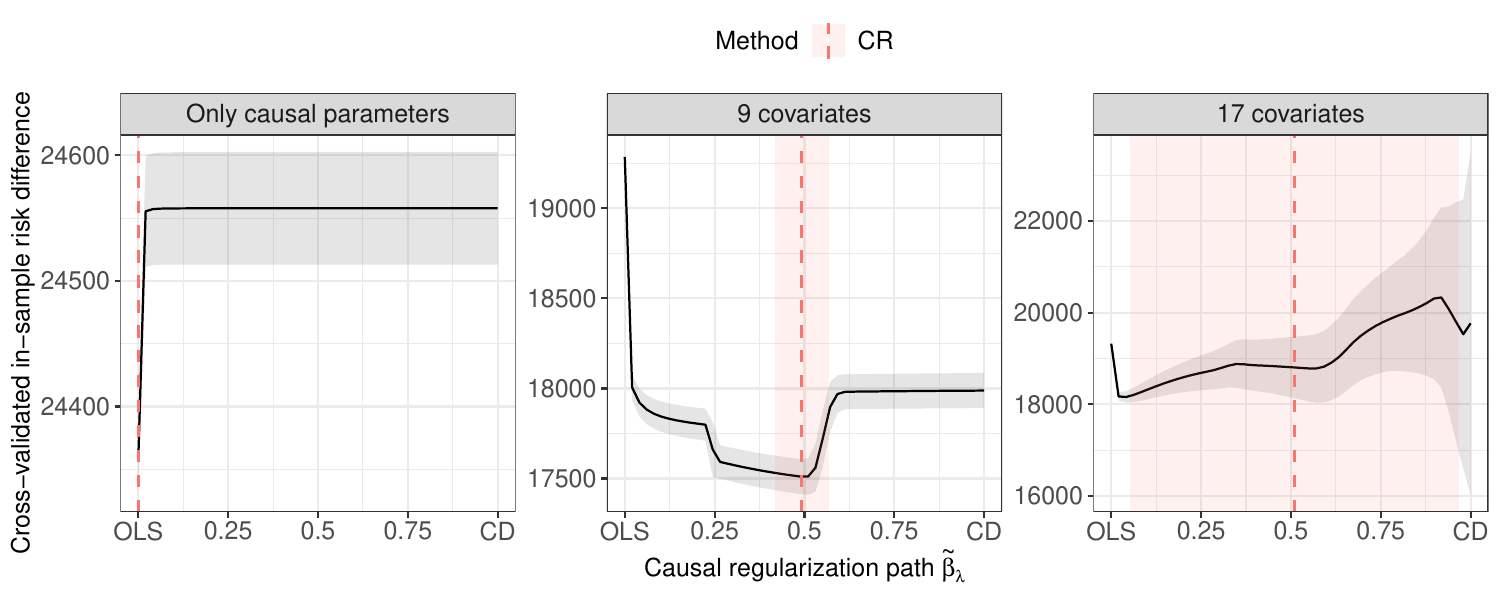}
\caption{Cross-validated in-sample risk difference for the light tunnel across three experiments. The red line indicates the average selected causal regularizer using 10-fold cross-validation. All shaded regions represent one standard deviation across $1000$ resamples. The causal regularization path has been normalized, see \eqref{eq:normalization}.}
\label{fig:chamber_in_sample_10}
\end{figure}

\begin{figure}[ht!]
\centering
\includegraphics[width=\linewidth]{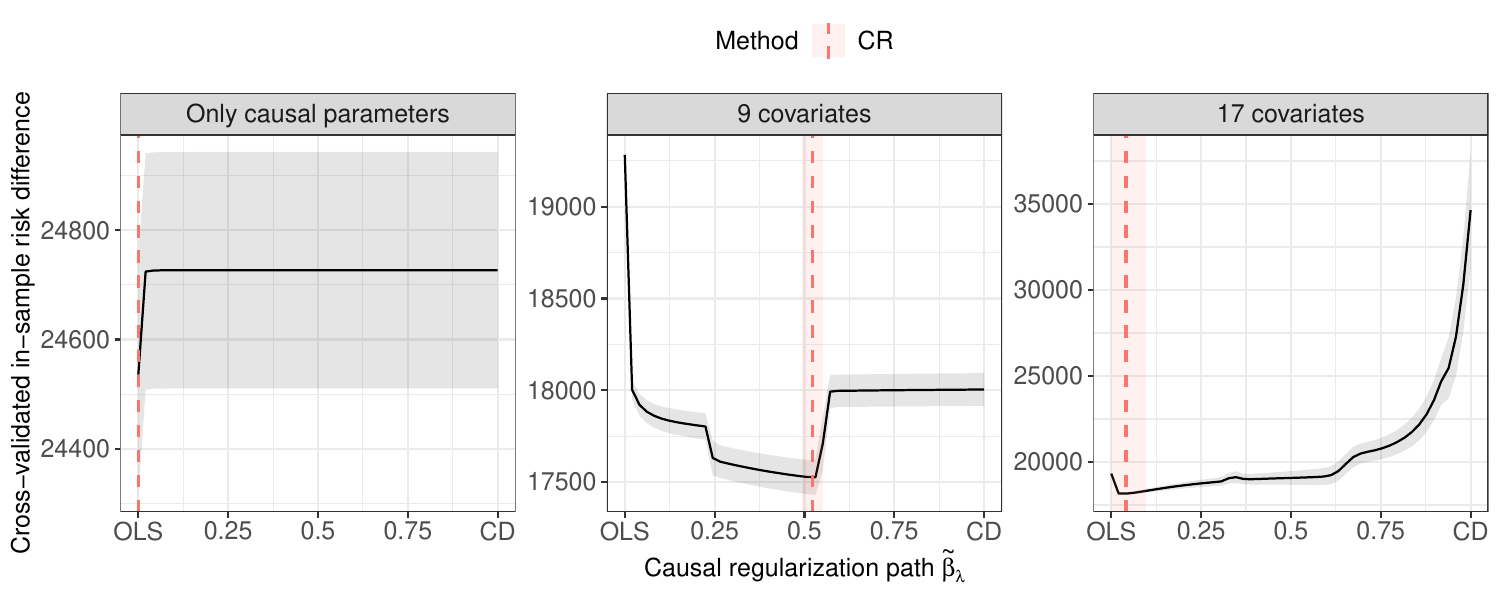}
\caption{Cross-validated in-sample risk difference for the light tunnel across three experiments. The red line indicates the average selected causal regularizer using 50-fold cross-validation. All shaded regions represent one standard deviation across $1000$ resamples. The causal regularization path has been normalized, see \eqref{eq:normalization}.}
\label{fig:chamber_in_sample_50}
\end{figure}

\zcref[S]{fig:chamber_in_sample_10} displays the in-sample absolute risk difference along the full causal regularization path, including OLS, the causal Dantzig, and the model selected via 10-fold cross-validation. In the first two panels, which correspond to the low-dimensional setting, 10-fold cross-validation selects well-performing models. In contrast, the third panel, representing the high-dimensional setting, reveals a substantial increase in variance. In practice, it is advised to increase the number of folds with the dimension to stabilize the selection procedure. \zcref[S]{fig:chamber_in_sample_50} shows the results of applying causal regularization with 50-fold cross-validation, which stabilizes the selection procedure across all settings.

\zcref[S]{fig:chamber_out_of_sample_50} presents the out-of-sample risk along the causal regularization path using 50-fold cross-validation. In the first panel, where only the causal parents are used, OLS and the causal Dantzig produce identical results, so all choices perform similarly. In the second panel, adding extra covariates introduces confounding, and causal regularization selects a model that improves out-of-sample performance over OLS. In the third panel, including all variables dilutes the in-sample distribution shift, and the selected causal regularizer closely matches OLS, which achieves the lowest out-of-sample risk.

\begin{figure}[ht!]
\centering
\includegraphics[width=\linewidth]{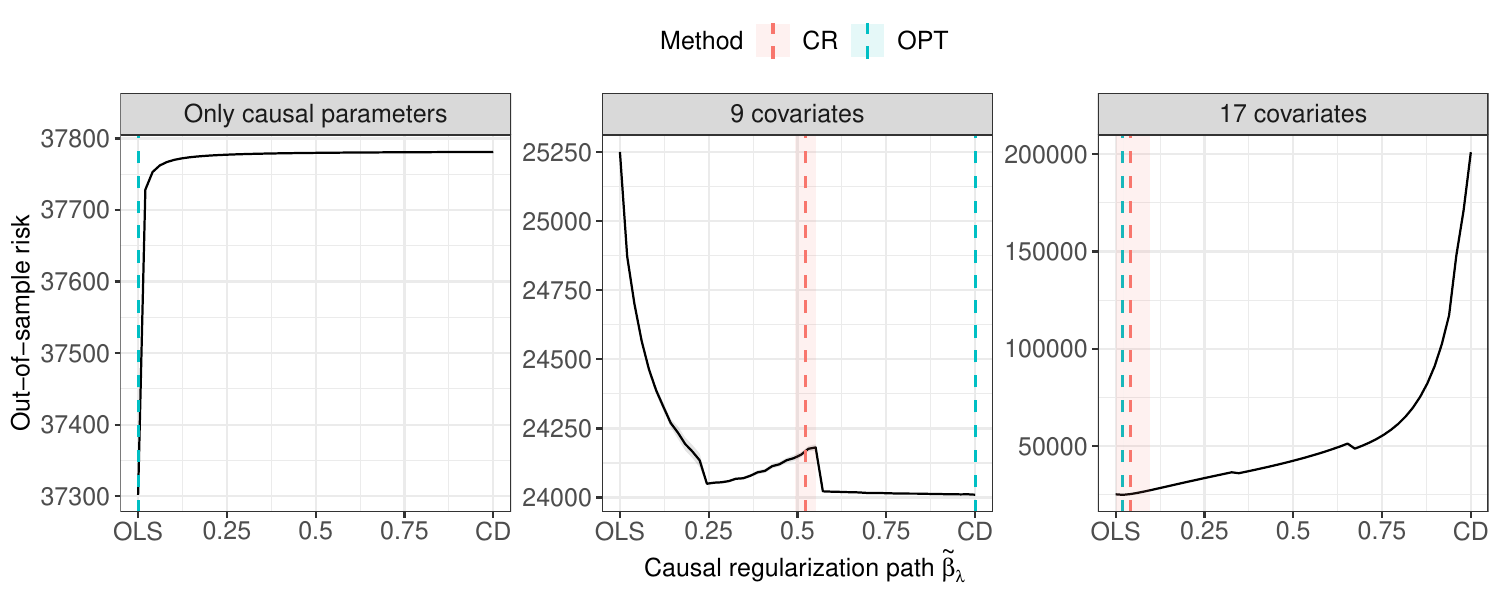}
\caption{Out-of-sample risk for the light tunnel across three experiments. The blue dashed line indicates the optimal $\gamma$ parameter for out-of-sample risk. The red dashed line indicates the average selected causal regularizer using 50-fold cross-validation. The blue dashed line indicates the optimal choice based on the test dataset. All shaded regions represent one standard deviation across $1000$ resamples. The causal regularization path has been normalized, see \eqref{eq:normalization}.}
\label{fig:chamber_out_of_sample_50}
\end{figure}

Additional experiments, in which the chosen causal regularizer closely resembles ordinary least squares due to the absence of in-sample distribution shift, have been deferred to \zcref{sec:additional_experiments}.

\section{Conclusion}

In this paper, we introduced \textit{causal regularization}, a technique for trading off in-sample and out-of-sample risk guarantees by leveraging heterogeneity across in-sample datasets. This method ensures out-of-sample guarantees for any regularization parameter against specific distribution shifts, similar to the causal Dantzig estimator, while remaining identifiable in a broader range of cases. We also showed, both theoretically and empirically, that cross-validation enables effective model selection for causal regularization, producing estimators with improved out-of-sample performance compared to OLS and the causal Dantzig.

Although causal regularization is designed to address additive distribution shifts, as defined in \zcref{def:sem}, when the shift deviates from this structure, the method does give risk guarantees for future environments in the span of its projection onto the space of additive shifts. Its predictive accuracy can improve further by learning a suitable basis from part of the data and using it to perform regression on the remainder.

The search for models that provide out-of-sample guarantees as a consequence of in-sample invariance parallels the idea of algorithmic stability \citep{surveystability}. The main difference is that the latter studies out-of-sample risk bounds, assuming that the training and test data come from the same distribution. Algorithmic stability requires \emph{model stability} \citep{modelstability}, that is, under sample perturbation, the algorithm must return similar models, in the sense that their predictions do not differ much on average.  \citet{riskstability-kearns} and \citet{riskstability-bousquet} relaxed the stringent requirement of model stability to the requirement of \emph{risk stability}, i.e., an algorithm returns models that have similar risk given perturbed samples. The work of \citet{icp} follows the model stability approach, where they look for a model with similar predictions under perturbed sub-samples. In contrast, \citet{anchorreg}, when using a categorical instrument that indicates the sub-samples, looks for a model that provides similar risk across sub-samples. Although the connection with risk stability is not immediate from their results, our work makes the requirement explicit and helps to elucidate the relationship between the two subfields.

The ideas presented here may extend to some generalised linear models by defining risks that promote stability under the causal parameters. For example, consider the Poisson regression analogue of \zcref{def:sem}: $\Ye|\Xe \sim \text{Poisson}(m(\betapa))$, where $m(\beta) \defeq \exp(\beta^T \Xe)$, and define the Pearson $\chi^2$-risk as $\Renv(\beta) \defeq \E\left[(\Ye - m(\beta))^2 / m(\beta)\right]$. This risk remains constant at $\betapa$ under perturbations of $A$. \citet{polinelliGeneralisedCausalDantzig2024} used this property to extend the causal Dantzig to Poisson models. When this generalised estimator has high variance or is not identifiable, we can bias it toward OLS using causal regularization, as in \eqref{def:causalregularization}, to trade off in-sample fit for stability across sub-samples. However, deriving the corresponding out-of-sample guarantees remains challenging due to the model’s non-linearity and requires additional assumptions.

\section*{Acknowledgment}
We thank Philip Kennerberg, Alice Polinelli, and Veronica Vinciotti for their valuable discussions during the development of this work. We also thank the reviewers of an earlier version of this work for their insightful comments, which significantly improved the paper. The authors gratefully acknowledge funding from the Swiss National Science Foundation (SNSF 188534).

\pagebreak

\appendix
\begin{center}

{\large\bf SUPPLEMENTARY MATERIAL}

\end{center}

\addcontentsline{toc}{section}{Supplementary material}
\etocdepthtag.toc{mtappendix}
\etocsettagdepth{mtchapter}{none}
\etocsettagdepth{mtappendix}{subsection}
\etocsettagdepth{mtreferences}{section}
\renewcommand{\contentsname}{Supplementary material}
{
\normalsize
\renewcommand{\contentsname}{\normalsize Table of contents}
\tableofcontents
}

\pagebreak

\section{Code to reproduce simulations and applications}

The code to reproduce the simulations and experiments can be accessed at \begin{center}
\url{https://github.com/lkania/Causal-Regularization}
\end{center} 

\section{Risk under structural equation model}\label{appx:proofs}

\begin{remark}[\bf Risk expansion under \zcref{def:sem}]Let $[Y_A,X_A]^T=\SEM(A)$ and $[Y_0,X_0)]^T=\SEM(0)$, we proceed to write some representations of the pooled risk and risk difference that are of use in the results throughout the appendix.

Under \zcref{def:sem}, both the covariates and the target can be written as projections of the sources of randomness \begin{align}
X_A &= u_x(\epsilon+A) \where u_x \in \R^{p\times(p+1)}\comma u_x = \PX(I-B)^{-1} \textand \PX = \begin{bmatrix}0_{p\times 1}& I_{p\times p}\end{bmatrix}\\
Y_A &= u_y(\epsilon+A) \where u_y \in \R^{1\times(p+1)}\comma u_y = \PY(I-B)^{-1} \textand \PY = \begin{bmatrix}
        1 & 0_{1\times p}
    \end{bmatrix}.
\end{align} It follows that the residuals of any linear model are \begin{align}
Y_A-\beta^{T}X_A = u_\beta(\epsilon+A) \where u_\beta \in \R^{1\times(p+1)} \textand  u_\beta=u_y-\beta^{T}u_x. \label{eq:residual}
\end{align} Recall that the risk $R_A$ is \begin{align}
&R_A(\beta) = \beta^{T}G_A\beta - 2\beta^{T}Z_A + W_A\\
&\where G_A = \E[X_AX_A^{T}]\comma Z_A = \E[X_AY_A] \textand W_A = \E[Y_A^2],
\end{align} and consequently the risk difference is \begin{align}
&R_\Delta(\beta) = R_A(\beta)-R_0(\beta)=\beta^{T}G_\Delta\beta - 2\beta^{T}Z_\Delta + W_\Delta\\
&\where G_\Delta = G_A-G_0\comma Z_\Delta = Z_A-Z_0 \textand W_\Delta = W_A-W_0.
\end{align} We expand each quantity, for the Gram matrix of $X_A$ it follows that \begin{align}
G_A &= \E[X_AX_A^{T}] = u_x\left(\E[\epsilon\epsilon^{T}]+\E[AA^{T}]\right)u_x^{T},\\
G_\Delta &= G_A - G_0 = u_x\left(\E[AA^{T}]\right)u_x^{T}, \label{eq:Gdelta}\\
\textand G_+ &= G_A + G_0 = u_x\left(2\E[\epsilon\epsilon^{T}]+\E[AA^{T}]\right)u_x^{T}.
\end{align} For the second moment between the covariates and the target, it follows that \begin{align}
Z_A &= \E[X_AY_A] = u_x\left(\E[\epsilon\epsilon^{T}]+\E[AA^{T}]\right)u_y^{T}= G_A\betapa + u_x\left(\E[\epsilon\epsilon_Y]+\E[AA_Y]\right),\\
Z_\Delta &= Z_A - Z_0 = u_x\E[AA^{T}]u_y^{T} = G_\Delta\betapa + u_x\E[AA_Y] \label{eq:Zdelta},\\
\textand Z_+ &= Z_A + Z_0 =  u_x\left(2\E[\epsilon\epsilon^{T}]+\E[AA^{T}]\right)u_y^{T} = G_+\betapa + u_x\left(2\E[\epsilon\epsilon_Y]+\E[AA_Y]\right).
\end{align} Finally, for the difference of second moments of the target, it follows that \begin{align}
W_A &= E[Y_A^2] = \betapat G_A\betapa + \E\epsilon_Y^2 + \E A_Y^2 + 2\betapat u_x\left(\E[\epsilon\epsilon_Y]+\E[AA_Y]\right),\\
W_\Delta &= W_A - W_0=\betapat G_\Delta\betapa + \E A_Y^2 + 2\betapat u_x\E[AA_Y],\\
\textand W_+ &= W_A + W_0 = \betapat G_+\betapa + 2\E\epsilon_Y^2 + \E A_Y^2 + 2\betapat u_x\left(2\E[\epsilon\epsilon_Y]+\E[AA_Y]\right).
\end{align} Putting it all together, we have that the risk $R_A$ can be represented as
\begin{align}
R_A(\beta) &= u_\beta \left(\E\left[\epsilon\epsilon^T\right] + \E\left[AA^{T}\right]\right)u_\beta^{T} \label{eq:risk_A}\\
\textand R_A(\beta) &= (\beta-\betapa)^{T}G_A(\beta-\betapa) - 2(\beta-\betapa)^{T}u_x\left(\E[\epsilon\epsilon_Y]+\E[AA_Y]\right) + \E A_Y^2 + \E\epsilon_Y^2.
\end{align} Consequently, we have the following representations of the risk difference \begin{align}
R_\Delta(\beta) &= u_\beta  \E\left[AA^{T}\right]u_\beta^{T} \label{eq:risk_diff}\\
\textand R_\Delta(\beta) &= (\beta-\betapa)^{T}G_\Delta(\beta-\betapa) - 2(\beta-\betapa)^{T}u_x\E[AA_Y] + \E A_Y^2.
\end{align} Finally, The pooled risk is \begin{align}
&R_+(\beta) = R_A(\beta)+R_0(\beta)=\beta^{T}G_+\beta - 2\beta^{T}Z_+ + W_+\\
&\where G_+ = G_A+G_0\comma Z_+ = Z_A+Z_0 \textand W_+ = W_A+W_0.
\end{align} Hence, it holds that for $u_\beta = u_y-\beta^T u_x$ \begin{align}
R_+(\beta) &= u_\beta \left(2\E\left[\epsilon\epsilon^T\right] + \E\left[AA^{T}\right]\right)u_\beta^{T} \label{eq:risk_OLS}\\
\textand R_+(\beta)&= (\beta-\betapa)^{T}G_+(\beta-\betapa) - 2(\beta-\betapa)^{T}u_x\left(2\E[\epsilon\epsilon_Y]+\E[AA_Y]\right) + \E A_Y^2 + 2\E\epsilon_Y^2.
\end{align}

\end{remark}

\begin{remark}[\bf Explicit formula for $u_x$] In the above remark, $u_x$ appears in all risk measures. It will be useful to have an explicit formula for it. Recall that \begin{equation}
u_x = \PX (I - B)^{-1}.
\end{equation} Let the following variables denote the inverse of $I-B$ in \zcref{def:sem} \begin{equation}
(I-B)^{-1} = \begin{bmatrix}
    1 & v^T\\
    w & M\\
\end{bmatrix}.
\end{equation} It follows that $u_x$ can be expressed as follows \begin{equation}
u_x = \begin{bmatrix}
    w & M
\end{bmatrix}.
\end{equation} We proceed to give explicit formulas for $M$ and $w$. From the invertibility of $(I-B)$, it follows that \begin{align}
    (I-B)^{-1}(I-B) = I_{(p+1)\times(p+1)} &\implies \begin{cases}
        -w\betapat+M(I-B_X) &= I_{p\times p}\\
        w-M\betach&=0_{p\times 1}
    \end{cases}.
\end{align} Consequently, \begin{equation}
M \left((I-\Bx)-\betach\betapat\right) = I_{p\times p},
\end{equation} which implies that $M$ has a right-inverse. Since $M$ is square, it is invertible and it follows that \begin{equation}
M  = \left((I-\Bx)-\betach\betapat\right)^{-1}. \label{eq:Mdefinition}
\end{equation} For $w$ it holds that \begin{equation}
 w=M\betach. \label{eq:wdefinition}
\end{equation} Putting it all together, we have that \begin{equation}\label{eq:ux_expansion}
u_x = \begin{bmatrix}
    M\betach & M
\end{bmatrix} = M \begin{bmatrix}
    \betach & I_{p\times p}
\end{bmatrix}.
\end{equation}
\end{remark}

\section{Causal Dantzig}\label{sec:causal_dantzig}

\begin{remark}\label{minimimum_cd}Let $V_A=[Y_A,X_A]^T$ and $V_0=[Y_0,X_0]^T$ be random variables whose second moments exist. Any $\beta$ that minimizes the absolute risk difference \begin{equation}
\beta \in \argmin_\beta |R_\Delta(\beta)|
\end{equation} must satisfy the condition \begin{equation}
0 \in (2G_\Delta\beta - 2Z_\Delta)\cdot \xi  \where \xi \in \begin{cases}
\left\{\sign(R_\Delta(\beta))\right\} &\textif R_\Delta(\beta)\not=0\\
[-1,1] &\textif  R_\Delta(\beta)=0
\end{cases}.
\end{equation} Consequently, it must hold that \begin{equation}
G_\Delta\beta = Z_\Delta \text{ and/or } R_\Delta(\beta)=0.
\end{equation}

If $R_\Delta(\beta)\geq 0$ for all $\beta \in \Rp$, then the condition simplifies to \begin{equation}
\beta \in \argmin_\beta |R_\Delta(\beta)| \iff G_\Delta \beta = Z_\Delta.
\end{equation}
\end{remark}

\begin{remark}\label{risk_diff_positivity} Let $(X_A,Y_A)=\SEM(A)$ and $(X_0,Y_0)=\SEM(0)$.
By \eqref{eq:risk_diff}, we have that \begin{equation}
R_\Delta(\beta) = u_\beta  \E\left[AA^{T}\right]u_\beta^{T}.
\end{equation} Since $\E\left[AA^{T}\right]$ is positive semi-definite, it follows that $R(\beta)\geq 0\quad \forall \beta \in \Rp$.
\end{remark}

\begin{restatable}{lemma}{MinimimumCDUnderSEM}\label{lemma:minimimum_cd_under_SEM}
Let $(X_A,Y_A)=\SEM(A)$ and $(X_0,Y_0)=\SEM(0)$. \begin{equation}
\betacd \in \argmin_{\beta\in\Rp} |R_\Delta(\beta)|.
\end{equation} It holds that \begin{equation}
\Gdelta \betacd = G_\Delta\betapa + M\left(\betach \E[A_Y^2] + \E[A_XA_Y]\right),
\end{equation} where $M=\left((I-\Bx)-\betach\betapat\right)^{-1}$.\end{restatable}
% \MinimimumCDUnderSEM*
\begin{proof}[Proof of \zcref{lemma:minimimum_cd_under_SEM}] \zcref{risk_diff_positivity} states that the risk difference $\Rdiff$ is a non-negative function. By \zcref{minimimum_cd}, it follows that $\betacd$ satisfies \begin{equation}
\Gdelta \betacd = \Zdelta.
\end{equation} By \eqref{eq:Zdelta} and \eqref{eq:ux_expansion}, it follows that \begin{equation}
\Zdelta = \GDelta\betapa + M\left(\betach \E[A_Y^2] + \E[A_XA_Y]\right),
\end{equation} from which the lemma follows.
\end{proof}

The following corollary is an extension of  \zcref{prop:causal_identification}.

\begin{corollary}[\bf Identification of causal parameters via causal Dantzig]\label{prop:causal_identification_extended} Assume that $(X_A,Y_A)=\SEM(A)$, $(X_0,Y_0)=\SEM(0)$, and that the target is not directly shifted, i.e. $A_Y = 0$, then it holds that any causal Dantzig solution satisfies \begin{equation}
\betacd \in \argmin_\beta |R_\Delta(\beta)| \iff G_\Delta\beta =G_\Delta\betapa.
\end{equation} If $\E\left[A_XA_X^{T}\right]$ is full-rank, then it follows that the causal Dantzig coincides with the causal parameters \begin{equation}
\betacd = \betapa.
\end{equation}\end{corollary}\begin{proof} By \zcref{lemma:minimimum_cd_under_SEM} and the fact that $A_Y=0$, it holds that \begin{equation}
\Gdelta \betacd = G_\Delta\betapa.
\end{equation} Additionally, by \eqref{eq:ux_expansion}, it follows that \begin{equation}
G_\Delta = u_x\begin{bmatrix}
    0 & 0_{1 \times p}\\
    0_{p\times 1} & \EAeAe
\end{bmatrix}u_x^{T} = M\EAeAe M^{T} \where M= \left((I-\Bx)-\betach\betapat\right)^{-1}.
\end{equation} Since $M$ is invertible, it is a product of elementary matrices and does not modify the column-rank or the row-rank of any conformal matrix. Consequently, \begin{equation}
\rank G_\Delta = \rank \EAeAe.
\end{equation} Thus if $\EAeAe$ is full-rank, it follows that $\Gdelta$ is full-rank, and consequently $\betapa = \betacd$.
\end{proof}

\section{Interpretation of regularizer}

\begin{proposition}
Assume that $[Y_A,X_A]^T=\SEM(A)$, $[Y_0,X_0]^T=\SEM(0)$, and $\E[\epsilon|A]=0$ in \zcref{def:sem} then \begin{equation}
R_\Delta(\beta)=\E[(\E[Y_A-X_A^{T}\beta|A])^2]
\end{equation}\end{proposition}\begin{proof}
Since $\E[\epsilon|A]=0$, by \eqref{eq:residual} it holds that \begin{equation}
\E[Y_A-\beta^{T}X_A|A]=u_\beta A \label{eq:projected_residual}
\end{equation} Finally, by \eqref{eq:risk_diff}, it follows that \begin{equation}
\E\left(\E[Y_A-\beta^{T}X_A|A]\right)^2=\E\left(u_\beta A\right)^2=u_\beta \E[AA^{T}]u_\beta^{T}=R_\Delta(\beta)
\end{equation}
\end{proof}

\begin{proposition}[\bf Correlation between shift and projected residuals] Assume that $[Y_A,X_A]^T=\SEM(A)$, $[Y_0,X_0]^T=\SEM(0)$, and $\E[\epsilon|A]=0$ in \zcref{def:sem} then
\begin{equation}
\norm{\Gdelta\beta-\Zdelta}_2^2 = \norm{\ u_x\ \E\left[\ A\ \E[Y_A-X_A^{T}\beta|A]\ \right]\ }_2^2
\end{equation}\end{proposition}\begin{proof}
The following sequence of equalities holds:
\begin{align}
\Gdelta\beta-\Zdelta &= u_x\E[AA^{T}]u_x^{T}\beta-u_x\E[AA^{T}]u_y\\
&=u_x\E[AA^{T}]u_\beta^{T}\\
&=u_x\E[A\left(u_\beta A\right)^{T}]\\
&=u_x\E[A\ \E\left[Y_A-X_A^{T}\beta|A\right]\ ] &&\by \eqref{eq:projected_residual}.\\
\end{align}
\end{proof}

\section{Causal Regularization}\label{appx:causal_regularization}

\begin{theorem}[\bf Moore–Penrose inverse]\label{thm:penrose}
Given a $A \in \R^{p\times p}$, the Moore–Penrose inverse $A$ is defined as the unique matrix $A^\dagger  \in \R^{p\times p}$ that satisfies the following axioms\begin{align}
A^\dagger  A A^\dagger  &= A, \label{eq:penrose_1}\\
A A^\dagger  A &= A, \label{eq:penrose_2}\\
(A A^\dagger )^T  &= A A^\dagger \textand  \label{eq:penrose_3}\\
(A^\dagger  A)^T  &= A^\dagger  A.  \label{eq:penrose_4}
\end{align} The following is a consequence of the above axioms: \begin{align}
\left(A^{T} A\right)^\dagger  A &= A^\dagger.  \label{eq:pseudoinverse_ls}
\end{align}

\end{theorem}

\subsection{Worst risk decomposition}\label{appx:decomposition}

The following lemma encompasses both \zcref{lemma:worst_risk_decomposition} and \zcref{lemma:control_via_grad}.

\begin{restatable}[\bf Worst risk decomposition]{lemma}{WorstRiskDecomposition}
\label{lemma:decomposition} Let $[Y_A,X_A]^T=\SEM(A)$ and $[Y_0,X_0]^T=\SEM(0)$ then \begin{equation}\label{eq:linear_risk_control}
\argmin_{\beta \in \mathbb{R}^p}\sup_{\tilde{A} \in \mathcal{C}_\lambda} R_{\tilde{A}}(\beta) = \argmin_{\beta \in \mathbb{R}^p}\frac{1}{2}R_+(\beta) +\frac{\lambda}{2}|R_\Delta(\beta)|,
\end{equation} where $\mathcal{C}_\lambda$ is the set of shift that are $\lambda$-times stronger than $A$ \begin{equation}
\mathcal{C}_\lambda = \left\{ \tilde{A} \in \mathcal{A} : \E[\tilde{A}\tilde{A}^T] \preceq \frac{1+\lambda}{2} \cdot \E[AA^T]  \right\},
\end{equation} and $N \preceq M$  if $M - N$ is positive semi-definite. Furthermore, there exists constants $K \geq 0$ and $r > 0$ such that \begin{equation}
\min_{\beta\in\Rp}\sup_{\tilde{A}\in\mathcal{C}_{\lambda}} R_{\tilde{A}}(\beta) \leq K \cdot \lambda + \min_{\beta\in\Rp} \frac{1}{2}R_+(\beta) + r \cdot \frac{\lambda}{2} \cdot \norm{G_\Delta\beta-Z_\Delta}_2^2.
\end{equation}
\end{restatable}
\begin{proof}
By \eqref{eq:risk_A} and definition of $\mathcal{C}_\lambda$, it holds that \begin{align}
\sup_{\tilde{A}\in \mathcal{C}_\lambda} R_{\tilde{A}}(\beta) &= \sup_{\tilde{A}\in \mathcal{C}_\lambda}u_\beta \left(\E\left[\epsilon\epsilon^T\right] + \E\left[\tilde{A}\tilde{A}^{T}\right]\right)u_\beta^{T}\\
&= u_\beta \left(\E\left[\epsilon\epsilon^T\right] + \frac{1+\lambda}{2}\cdot \E\left[AA^{T}\right]\right)u_\beta^{T}.
\end{align} Using \eqref{eq:risk_OLS} and \eqref{eq:risk_diff}, it follows that \begin{equation}
\sup_{\tilde{A}\in\mathcal{C}_\lambda} R_{\tilde{A}}(\beta) = \frac{1}{2}R_+(\beta) + \frac{\lambda}{2}\cdot R_\Delta(\beta).
\end{equation} By \zcref{risk_diff_positivity} $\Rdiff$ is a non-negative function, and the first claim of the lemma follows.

For the second claim, we split the analysis into two cases. If $\rank G_\Delta >0$, then let $\bar{\beta} \in \Rp$ be any vector that satisfies $\GDelta \bar{\beta} = \Zdelta$, it follows that \begin{align}
R_\Delta(\beta) &= R_\Delta(\bar{\beta}) + (\GDelta \bar{\beta} -\Zdelta)^{T}(\beta-\betacd) + (\beta-\bar{\beta})^{T}\Gdelta(\beta-\bar{\beta})\\
&= R_\Delta(\bar{\beta}) + (\beta-\bar{\beta})^{T}\Gdelta(\beta-\bar{\beta}).
\end{align} Since $\Gdelta$ is positive semi-definite, it follows that there exists a positive semi-definite matrix $G_\Delta^{1/2}$ that is its square root \begin{equation}
G_\Delta = \left(G_{\Delta}^{1/2}\right)^{T}G_{\Delta}^{1/2}
\end{equation} Thus, it follows that \begin{equation}
(\beta-\bar{\beta})^{T}\Gdelta(\beta-\bar{\beta}) = \norm{G_\Delta^{1/2}(\beta-\bar{\beta})}_2^2.
\end{equation} Let $\Gsrg$ be the pseudoinverse of $\Gsr$, note that
\begin{align}
    \Gsr &= \Gsr\Gsrg\Gsr &&\by \eqref{eq:penrose_2}\\
    &= (\Gsr\Gsrg)^{T}\Gsr &&\by \eqref{eq:penrose_3}\\
    &= (\Gsrg)^{T}(\Gsr)^{T}\Gsr\\
    &= (\Gsrg)^{T}\Gdiff. &&\text{by the definition of square root}
\end{align}	Consequently, \begin{align}
R_\Delta(\beta) &= R_\Delta(\bar{\beta}) + \norm{G_\Delta^{g/2}G_\Delta(\beta-\bar{\beta})}_2^2\\
&=R_\Delta(\bar{\beta}) + \norm{G_\Delta^{g/2}\left(G_\Delta\beta-Z_\Delta\right)}_2^2 &&\since G_\Delta\bar{\beta}=Z_\Delta\\
&\leq R_\Delta(\bar{\beta}) + \norm{G_\Delta^{g/2}}^2 \cdot \norm{G_\Delta\beta-Z_\Delta}_2^2
\end{align} where $\norm{\cdot}$ is the operator norm, see \zcref{def:operator_norm}. It follows that \begin{equation}
\min_{\beta\in\Rp}\sup_{\tilde{A}\in\mathcal{C}_{\lambda}} R_{\tilde{A}}(\beta) \leq K \cdot \lambda + \min_{\beta\in\Rp} \frac{1}{2}R_+(\beta) + r \cdot \frac{\lambda}{2} \cdot \norm{G_\Delta\beta-Z_\Delta}_2^2,
\end{equation} where \begin{equation}
K = \frac{1}{2}\cdot\argmin_{\beta : \Gdelta\beta=\Zdelta}|R_{\Delta}(\beta)| \textand r = \sigma^{-1}
\end{equation} and $\sigma$ is the smallest non-vanishing singular value \begin{equation}
\sigma = \norm{G_\Delta^{g/2}}^{-2} = \min_{1\leq i \leq p}\left\{ \left[\sigma_i(G_\Delta^{1/2})\right]^2 : \sigma_i(G_\Delta^{1/2}) \not= 0 \right\}.
\end{equation} Finally, if $\rank G_\Delta =0$, then $G_\Delta = \bzero$, which implies that \begin{equation}
0 = \beta^{T}G_\Delta\beta =\beta^{T}u_x\E[AA^{T}]u_x^{T}\beta \norm{\E[AA^{T}]^{1/2}u_x^{T}\beta}_2^2 \iff \E[AA^{T}]^{1/2}u_x^{T}\beta=0 \forall \beta.
\end{equation} Consequently, it follows that \begin{equation}
\beta^{T}Z_\Delta = \beta^{T}u_x\E[AA^{T}]u_y^{T} = \left(\E[AA^{T}]^{1/2}u_x^{T}\beta\right)^{T}\E[AA^{T}]^{1/2} =0\ \forall \beta.
\end{equation}

Thus $
|R_\Delta(\beta)| = |W_\Delta|
$ and it follows that \begin{equation}
|R_\Delta(\beta)| \leq |W_\Delta| + \frac{\lambda}{2} \cdot \norm{G_\Delta\beta-Z_\Delta}_2^2 \quad \forall \beta,
\end{equation} which proves that the lemma statement holds for $\rank G_\Delta =0$.

\end{proof}

\subsection{Consistency}\label{appx:consistency}

\begin{definition}[\bf Operator norm]\label{def:operator_norm}
Let $A \in \R^{p\times p}$ be a symmetric matrix, its operator norm can be defined as \begin{equation}
    \norm{A} = \max_{x\in\Rp \st \norm{x}_2\leq 1}|x^TAx|
\end{equation} and it holds that \begin{equation}
    \norm{A} = \max_{1\leq i\leq p}|\gamma_i(A)|
\end{equation} where $\gamma_i(A)$ is the $i$-th eigenvalue of $A$.
\end{definition}

\begin{corollary}[\bf Operator norm of pseudo-inverse]\label{eq:operator_norm_pseudoinverse} Let $A \in \R^{p\times p}$ be a symmetric matrix, \begin{equation}
\norm{A^{\dagger}} = \begin{cases} 0 &\textif \gamma_i(A) = 0\quad \forall 1\leq i \leq p \\
    \max \{\ |\gamma_i(A)|^{-1} : |\gamma_i(A)| > 0\ \}\ & \otherwise\end{cases}.
\end{equation} Thus, if $A$ is symmetric positive definite, it holds that \begin{equation}
\norm{A^{\dagger}} = \norm{A^{-1}} = \gamma_{\min}^{-1}(A).
\end{equation}
\end{corollary}\begin{proof}
Let $U D V^T$ be the SVD decomposition of $A$, it holds that \begin{equation}
    A^\dagger  = U D^\dagger  V^T \where \gamma_i(D^\dagger ) = \begin{cases}
    0 &\text{ if } \gamma_i(A) = 0\\
    |\gamma_i(A)|^{-1} &\text{ otherwise }
    \end{cases},
\end{equation} and we rewrite the operator norm of $A$ as \begin{equation}
\norm{A^\dagger } = \norm{U D^\dagger  V^T} = \norm{D^\dagger } = \begin{cases} 0 &\text{ if } \gamma_i(A) = 0\ \forall 1\leq i \leq p \\
\max \{\ |\gamma_i(A)|^{-1} : |\gamma_i(A)| > 0\ \}\ & \text{otherwise}\end{cases}
\end{equation} where the second equality is due to the operator norm being unitarily invariant.
\end{proof}

\begin{theorem}[Theorem 3.3 of \citet{pseudoinverse}]\label{thm:pseudoinverse}
Let $A \in \R^{p\times p}$ and $B \in \R^{p\times p}$, then \begin{equation}
\norm{A^\dagger -B^\dagger } \leq C \cdot \max\{\norm{A^\dagger }^2,\norm{B^\dagger }^2\} \cdot \norm{A-B}
\end{equation} where $\norm{\cdot}$ denotes the operator norm, and $C=(1+\sqrt{5})/2$.
\end{theorem}

\begin{corollary}\label{cor:concentration_pseudoinverse} Let $A \in \R^{p\times p}$ be symmetric positive definite and $B \in \R^{p\times p}$ symmetric. If \begin{equation}\label{eq:assumption}
\gamma_{\min}(A) \geq 2 \cdot \norm{A-B}\comma
\end{equation} it holds that \begin{equation}
\norm{A^\dagger -B^\dagger } \leq \frac{4C}{\gamma_{\min}^2(A)} \cdot \norm{A-B}.
\end{equation}
\end{corollary}\begin{proof}
Since $A$ is a positive symmetric definite matrix, it has an inverse, and it follows that \begin{equation}\label{eq:lb_operator_A}
\norm{A^\dagger } = \norm{A^{-1}} = \gamma_{\min}(A)^{-1}\period
\end{equation} Furthermore, Weyl's inequality states that if $B$ is close to $A$ in operator norm, then their eigenvalues are close, \begin{equation}
|\ \gamma_i(A)-|\gamma_i(B)|\ |\leq|\gamma_i(A)-\gamma_i(B)|\leq\norm{A-B} \quad \for 1\leq i\leq p\period
\end{equation} Thus, by the assumption \eqref{eq:assumption}, a lower bound for $|\gamma_i(B)| $ is \begin{equation}
\frac{\gamma_{\min}(A)}{2} \leq \gamma_{\min}(A) - \norm{A-B} \leq |\gamma_i(B)| \quad \for 1\leq i\leq p\period
\end{equation} Consequently by \zcref{eq:operator_norm_pseudoinverse}, it holds that \begin{equation}\label{eq:lb_operator_B}
\norm{B^\dagger} \geq \frac{2}{\gamma_{\min}(A)} \period
\end{equation}Thus, by \zcref{thm:pseudoinverse}, \eqref{eq:lb_operator_A} and \eqref{eq:lb_operator_B}, we have that \begin{equation}
\norm{A^\dagger -B^\dagger } \leq C \cdot \max\left\{\frac{1}{\gamma^2_{\min}(A)},\frac{4}{\gamma^2_{\min}(A)}\right\} \cdot \norm{A-B}^2\comma
\end{equation} which proves the statement.
\end{proof}

\begin{lemma}\label{lemma:pseudoinverse}
Let $A \in \R^{p\times p}$ be a positive symmetric define matrix, and $\{B_n\}_{n=1}^{\infty} \in \R^{p\times p}$ a sequence of symmetric matrices such that $\norm{A-B_n} \toP 0$, then $\norm{A^\dagger -B_n^\dagger } \toP 0$
\end{lemma}
\begin{proof} Define the event $
    E_n\ \defeq\ \{\gamma_{\min}(A) \geq 2 \cdot \norm{A-B_n}\}.
$ By \zcref{cor:concentration_pseudoinverse}, it follows that on the event $E_n$, the event $C_n$ happens \begin{align}
C_n = \left\{\norm{A^\dagger -B_n^\dagger } \leq \frac{4C}{\gamma^2_{\min}(A)} \cdot  \norm{A-B_n} \right\}.\label{eq:pseudoinverse_upperbound}
\end{align} Thus, for $ \epsilon > 0 \textand \delta > 0$ \begin{align}
&P(\norm{A^\dagger -B_n^\dagger }\geq \epsilon)\\
&= P(\ \left\{\norm{A^\dagger -B_n^\dagger }\geq \epsilon\right\} \cap E_n\ ) + P(\ \left\{\norm{A^\dagger -B_n^\dagger }\geq \epsilon\right\} \cap E_n^c\ )\\
&\leq P( C_n ) + P( E_n^c )
\end{align} Where in the last step, we used \eqref{eq:pseudoinverse_upperbound}. Since $ \norm{A-B_n}\toP0 $ and $\gamma_{\min}(A)>0$, we can choose $N\defeq N(\epsilon,\gamma_{\min}(A),\delta) \in \mathbb{N}$ such that $P(C_n) < \frac{\delta}{2}$ and $P(E_n^c) < \frac{\delta}{2}$ for all $n \geq N$, which implies that $P(\norm{A^\dagger -B_n^\dagger }\geq \epsilon) < \delta$ for all $n \geq N$ and consequently $\norm{A^\dagger -B_n^\dagger }\toP 0$.

\end{proof}

\consistency*
\begin{proof} By the weak law of large numbers and continuity, $\bGplus$, $\bG$, $\bZplus$, and $\bZ$ converge in probability to $\Gplus$, $\Gdiff$, $\Zplus$, and $\Zdiff$ correspondingly. Hence, by continuity, $\bGlambda$ and $\bZlambda$ converge in probability to $\Glambda$ and $\Zlambda$. Since $p$ is fixed, $\bGlambda \toP \Glambda$ implies $\norm{\Glambda - \bGlambda} \toP 0$. By assumption, $\Glambda$ is a symmetric positive definite matrix. Thus, \ref{thm:pseudoinverse} implies that $\norm{\gGlambda - \gbGlambda} \toP 0$, and consequently $\gbGlambda \toP \gGlambda$. By continuity, we have that \begin{equation}
\hat{\beta}_\lambda=\gbGlambda\bZlambda \toP \gGlambda\Zlambda=\beta_\lambda,
\end{equation} which finishes the proof.
\end{proof}

\subsection{Concentration}\label{appx:concentration}

The following concentration results are well-known and can be derived from the results in chapter 6 of \cite{wainwrightHighDimensionalStatisticsNonAsymptotic2019}.

\begin{proposition}\label{prop:subgaussian_concentration}
Let $V_A = [Y_A,X_A]^T \in \SG(\sigma^2)$. There exists positive constant $C_0$ and $C_1$ such that \begin{equation}
\max\left( \norm{\hat{G}_A-G_A}\ ,\ \norm{\hat{Z}_A-Z_A}_2 \right) \leq C_0 \cdot \sigma \cdot \max\left(\ \sqrt{\frac{p+\log1/\delta}{n}}\ ,\ \frac{p+\log1/\delta}{n} \right)
\end{equation} with probability at least $1-\delta$ and \begin{equation}
\max\left( \norm{\hat{G}_A-G_A}_\infty\ ,\ \norm{\hat{Z}_A-Z_A}_\infty\ ,\ \norm{\hat{W}_A-W_A}_\infty \right) \leq C_1 \cdot \sigma \cdot \sqrt{\frac{\log p + \log 1/\delta}{n}}
\end{equation} with probability at least $1-\delta$.
\end{proposition}

\begin{proposition}\label{prop:beta_concentration} Let $V_A=[Y_A,X_A]^T \in \SG(\sigma^2)$, $V_0=[Y_0,X_0]^T \in \SG(\sigma^2)$ and $n=\min(n_A,n_0)$. If $\gamma_{\min}(G_\lambda)>0$, there exists positive constant $C$ such that whenever \begin{equation}
n \geq \left[\max\left(\frac{2C\sigma}{\gamma_{\min}(G_\lambda)},C\sigma,1\right)\right]^2 \cdot (p+\log(1/\delta)).
\end{equation} It holds that \begin{equation}
\norm{\hat{\beta}_\lambda-\beta_\lambda}_2 \leq C' \cdot \left[\max\left(\frac{1}{\gamma_{\min}(G_\lambda)},1\right)\right]^2 \cdot\sigma \cdot \sqrt{\frac{p+\log(1/\delta)}{n}}
\end{equation} with probability at least $1-\delta$, where $C'$ is a positive constant that depends on $\lambda$, $\norm{Z_\lambda}_2$, $\norm{G_\Delta}$ and $\norm{Z_\Delta}_2$.
\end{proposition}\begin{proof}
Let \begin{equation}
g = \norm{\hat{G}_A-G_A}+\norm{\hat{G}_0-G_0} \textand z = \norm{\hat{Z}_A-Z_A}_2+\norm{\hat{Z}_0-Z_0}_2.
\end{equation}

\underline{\textbf{Upper bound for $\norm{\hat{\beta}_\lambda-\beta_\lambda}_2$}} Recall that
\begin{equation}
\norm{\hat{\beta}_\lambda-\beta_\lambda}_2=\norm{\hat{G}_\lambda^\dagger\hat{Z}_\lambda-G_\lambda^{\dagger}Z_\lambda}_2
\end{equation} It follows that \begin{equation}
\norm{\hat{\beta}_\lambda-\beta_\lambda}_2 \leq R_1 \cdot R_2 + \norm{G_\lambda^{\dagger}}\cdot R_2 + \norm{Z_\lambda}_2\cdot R_1
\end{equation} where \begin{equation}
R_1 = \norm{\hat{G}_\lambda^\dagger-G_\lambda^\dagger} \textand R_2 = \norm{\hat{Z}_\lambda-Z_\lambda}_2.
\end{equation}

\underline{\textbf{Upper bound for $R_2$}} \begin{equation}
R_2 \leq  R_3 + \lambda \cdot R_4 \where R_3=\norm{\hat{Z}_+-Z_+} \textand R_4=\norm{\hat{G}_\Delta \hat{Z}_\Delta-G_\Delta Z_\Delta}\period
\end{equation} For $R_4$, we have that  \begin{equation}
R_4 \leq R_5 \cdot R_6 + \norm{Z_\Delta}_2 \cdot R_5 + \norm{G_\Delta}\cdot R_6
\end{equation} where \begin{equation}
R_5 = \norm{\hat{G}_\Delta-G_\Delta} \textand R_6 = \norm{\hat{Z}_\Delta-Z_\Delta}_2\period
\end{equation} Finally, \begin{equation}
R_5 \leq g \textand \max( R_3\ ,\ R_6) \leq z.
\end{equation}

\underline{\textbf{Upper bound for $R_1$}} If \begin{equation}\label{eq:condition_spectral_concentration}
\gamma_{\min}(G_\lambda) \geq 2 \cdot  R_7 \where R_7 = \norm{\hat{G}_\lambda-G_\lambda},
\end{equation} by \zcref{lemma:pseudoinverse}, we have that \begin{equation}
R_1 \leq  \frac{4C}{\gamma^2_{\min}(G_\lambda)} \cdot R_7 \where R_7 = \norm{\hat{G}_\lambda-G_\lambda}\period
\end{equation} Then \begin{equation}
R_7 \leq R_8 + \lambda \cdot R_9 \where R_8 = \norm{\hat{G}_+-G_+} \textand R_9= \norm{\hat{G}_\Delta\hat{G}_\Delta-G_\Delta G_\Delta}\comma
\end{equation} and \begin{equation}
R_9 \leq R_5^2 + 2\norm{G_\Delta}\cdot R_5 \comma R_5 \leq g \textand R_8 \leq g\period
\end{equation}

\underline{\textbf{Concentration}} By union bound and \zcref{prop:subgaussian_concentration}, there exists positive constant $C$ such that \begin{equation}
\max(z,g) \leq \phi \where \phi = C \cdot \sigma \cdot \sqrt{\frac{p+\log(1/\delta)}{n}}
\end{equation} with probability at least $1-\delta$.

Let \begin{equation}
n \geq \left[\max\left(\frac{2C\sigma}{\gamma_{\min}(G_\lambda)},C\sigma,1\right)\right]^2 \cdot (p+\log(1/\delta))
\end{equation} Then with probability $1-\delta$, \eqref{eq:condition_spectral_concentration} holds and $\phi < 1$. Consequently, we have that \begin{equation}
R_1 \leq C_1\cdot \phi \where C_1=\frac{4C}{\gamma^2_{\min}(G_\lambda)}\cdot [(1+2\norm{G_\Delta})\lambda+1]
\end{equation} and \begin{equation}
R_2 \leq C_2\cdot \phi \where C_2=[(1+\norm{G_\Delta}+\norm{Z_\Delta}_2)\lambda+1]
\end{equation} Consequently, \begin{equation}
\norm{\hat{\beta}_\lambda-\beta_\lambda}_2 \leq C_3 \cdot \phi \where C_3=\left(C_1C_2 + \norm{G_\lambda^{\dagger}}C_2 + \norm{Z_\lambda}_2C_1\right)
\end{equation} holds with probability at least $1-\delta$. Note that there exists constant $C'$ depending on depends on $\lambda$, $\norm{Z_\lambda}_2$, $\norm{G_\Delta}$ and $\norm{Z_\Delta}_2$ such that \begin{equation}
C_3 \leq C' \cdot \left[\max\left(\frac{1}{\gamma_{\min}(G_\lambda)},1\right)\right]^2,
\end{equation} which proves the statement.
\end{proof}

\begin{remark}\label{prop:loss_concentration_via_beta_concentration} Let $
\ell(\beta) = \beta^{T}G\beta - 2\beta^{T}Z + W$. For any $G$ it holds that \begin{equation}
|\ell(\beta)-\ell(\tilde{\beta})|\leq M \cdot \max\left(\ \norm{\beta-\tilde{\beta}}_1^2\ ,\ \norm{\beta-\tilde{\beta}}_2\ \right)
\end{equation} where $M = \max\left(\ \norm{G}_\infty ,\ \norm{Z}_2\ \right)$. If $G$ is symmetric positive semi-definite, it holds that \begin{equation}
|\ell(\beta)-\ell(\tilde{\beta})|\leq M \cdot \max\left(\ \norm{\beta-\tilde{\beta}}_2^2\ ,\ \norm{\beta-\tilde{\beta}}_2\ \right)
\end{equation} where $M = \max\left(\ \norm{G} ,\ \norm{Z}_2\ \right)\period$
\end{remark}

\Concentration*
\begin{proof}

It holds that  \begin{equation}
2\cdot \ell_\lambda(\beta) = \beta^{T}G_\lambda\beta -2\beta^{T}Z_\lambda + (W_+ + \lambda Z_\Delta^{T}Z_\Delta)
\end{equation}  where $G_\lambda$ is symmetric positive semi-definite. Thus, by \zcref{prop:loss_concentration_via_beta_concentration}, we have that \begin{equation}
\ell_\lambda(\hat{\beta}_\lambda) \leq \inf_{\beta \in \Rp}\ell_\lambda(\beta) + \frac{1}{2}\cdot \max\left(\ \norm{G_\lambda} ,\ \norm{Z_\lambda}_2\ \right) \cdot \max\left(\ \norm{\beta-\tilde{\beta}}_2^2\ ,\ \norm{\beta-\tilde{\beta}}_2\ \right) \period
\end{equation} The proof follows by \zcref{prop:beta_concentration}.

\end{proof}

\subsection{\texorpdfstring{$\ell_1$}{l1} constraint}\label{appx:l1_penalty}

Consider causal regularization but where we use an $\ell_1$ constraint rather than minimum $\ell_2$ norm: \begin{equation}
\hat{\beta}_{\lambda,\ell_1} = \argmin_{\norm{\beta}_1 \leq L} \ell_\lambda(\beta).
\end{equation} Then following \cite{greenshteinPersistenceHighdimensionalLinear2004}, since the $\ell_1$ norm is always guarantee to be upper-bounded by $L$, and $\norm{\hat{G}_\lambda-G_\lambda}_\infty$ converges at a rate of $\frac{\log p}{n}$, it is easy to see that \begin{equation}\label{eq:l1_concetration}
\ell_\lambda(\hat{\beta}_{\lambda,\ell_1}) \leq \argmin_{\norm{\beta}_1 \leq L}\ell_\lambda(\beta) + C \cdot L^2 \cdot \sqrt{\frac{\log p + \log 1/\delta}{n}}
\end{equation} with probability at least $1-\delta$. We prove \eqref{eq:l1_concetration} in \zcref{prop:l1_contration}. However, we first prove an auxiliary result.

\begin{proposition}\label{prop:loss_concentration_by_bounded_norm}
Let \begin{equation}
\ell(\beta) = \beta^{T}G\beta - 2\beta^{T}Z + W
\end{equation} and \begin{equation}
\hat{\ell}(\beta) = \beta^{T}\hat{G}\beta - 2\beta^{T}\hat{Z} + \hat{W}\period
\end{equation} Then \begin{equation}
|\ell(\beta)-\hat{\ell}(\beta)|\leq M \cdot \left(\norm{\beta}_1+1\right)^2
\end{equation} where \begin{equation}
M = \max\left(\ \norm{A-\hat{A}}_\infty\ ,\ \norm{b-\hat{b}}_\infty\ ,\ \norm{W-\hat{W}}_\infty\ \right)\period
\end{equation}
\end{proposition}\begin{proof}
Let $\gamma=\begin{bmatrix}-1\\\beta\end{bmatrix}$, then it follows that \begin{equation}
\ell(\beta) = v^{T}\Sigma v \quad \where \Sigma = \begin{bmatrix} W & Z^T\\
Z & A
\end{bmatrix}\period
\end{equation} Thus, we have that \begin{equation}
|\ell(\beta)-\hat{\ell}(\beta)|=|v^T(\Sigma-\hat{\Sigma})v|\leq \norm{\Sigma-\hat{\Sigma}}_\infty \cdot \norm{v}_1^2.
\end{equation} Noting that \begin{equation}
\norm{v}_1 = \norm{\beta}_1 + 1 \textand \norm{\Sigma-\hat{\Sigma}}_\infty\leq M
\end{equation} completes the proof.
\end{proof}

\begin{proposition}\label{prop:l1_contration}Let $V_A=[Y_A,X_A]^T\in \SG(\sigma^2)$, $V_0=[Y_0,X_0]^T \in \SG(\sigma^2)$ and $n=\min(n_A,n_0)$. There exists a constant $C$ such that
\begin{equation}
\ell_\lambda(\hat{\beta}_{\lambda,\ell_1}) \leq \argmin_{\norm{\beta}_1 \leq L}\ell_\lambda(\beta) + C \cdot L^2 \cdot \sqrt{\frac{\log p + \log 1/\delta}{n}}
\end{equation} with probability at least $1-\delta$.\end{proposition}\begin{proof}
It holds that \begin{align}
2\cdot \ell_\lambda(\beta) &= \beta^{T}G_\lambda\beta -2\beta^{T}Z_\lambda + W_\lambda \where W_\lambda=(W_+ + \lambda Z_\Delta^{T}Z_\Delta)\\
2\cdot \hat{\ell}_\lambda(\beta) &= \beta^{T}\hat{G}_\lambda\beta -2\beta^{T}\hat{Z}_\lambda + \hat{W}_\lambda \where \hat{W}_\lambda=(\hat{W}_+ + \lambda \hat{Z}_\Delta^{T}\hat{Z}_\Delta)
\end{align} Thus, by \zcref{prop:loss_concentration_by_bounded_norm}, we have that \begin{align}
|\ell_\lambda(\beta)-\ell_\lambda(\beta)|\leq M \cdot (\norm{\beta}_1+1)^2 \label{eq:concentration}
\end{align} where \begin{equation}
M = \max\left(\ \norm{G_\lambda-\hat{G}_\lambda}_\infty\ ,\ \norm{Z_\lambda-\hat{Z}_\lambda}_\infty\ ,\ \norm{W_\lambda - \hat{W}_\lambda}_\infty\ \right)\period
\end{equation} Then we have that \begin{align}
\ell_\lambda(\hat{\beta}_{\lambda,\ell_1}) &\leq \hat{\ell}_\lambda(\hat{\beta}_{\lambda,\ell_1}) + M \cdot (L+1)^2 &&\by \eqref{eq:concentration}\\
&\leq \argmin_{\norm{\beta}_1\leq L}\hat{\ell}(\beta) + M \cdot (L+1)^2 &&\text{by optimality of } \hat{\beta}_{\lambda,\ell_1}\\
&\leq \argmin_{\norm{\beta}_1\leq L}\ell(\beta) + 2M \cdot (L+1)^2 &&\by \eqref{eq:concentration}.
\end{align} Finally, using the inequalities in the proof of \zcref{prop:beta_concentration}, there exists a constant $C'$ that depends on $\lambda$, $\norm{Z_\Delta}_2$, $\norm{G_\Delta}$ such that \begin{equation}
M \leq C' \cdot \max(M_A,M_0)
\end{equation} where $M_A= \max\left(\ \norm{G_A-\hat{G}_A}_\infty\ ,\ \norm{Z_A-\hat{Z}_A}_\infty\ ,\ \norm{W_A - \hat{W}_A}_\infty\ \right)$. Thus, by union bound and \zcref{prop:subgaussian_concentration}, we have that there exists constant $C$ that depends on $C'$ such that \begin{equation}
M \leq C \cdot \sigma \cdot \sqrt{\frac{\log p + \log 1/\delta}{n}}
\end{equation} with probability at least $1-\delta$.

\end{proof}

\subsection{Model selection}\label{appx:model_selection}

We consider selecting the regularization parameter based on resampling. The main idea is to design a model selection procedure that asymptotically chooses the same model as an oracle would with access to the unknown distributions $(\Xo,\Yo)$ and $(\Xe,\Ye)$. The results of this section follow from the work of \citet{cvvanderlaan}. 

Given a set of models $\{\hat{\beta}_{\lambda}: \lambda \in \Lambda\}$ fitted on the in-sample data, a selector $\lambda$ is a choice of one of the models, i.e., $\lambda\in \Lambda$. Two selectors $\hat{\lambda}$ and $\lambda$ are \emph{asymptotically equivalent} under some loss $\ell$, if their difference vanishes in probability asymptotically, \begin{equation}
    E|\ell(\hat{\lambda}) - \ell(\lambda)| \to 0 \as n\to\infty
\end{equation}

We formally define data resampling in order to later recover the cross-validation and sample-splitting losses. Let $S^A = (S^A_1,\dots, S^A_{n_A}) \in \{0,1\}^{n_A}$ be a random variable that indicates if an observation from the dataset $(\bXep,\bYep)$ is in the train or test set. That is, $S^A_i = 0$ means that the $i$th observation is in the training set, while $S^A_i = 1$ indicates that it belongs to the test set. The distribution of $S^A$ determines the methodology. For instance, if its distribution is a point mass at a unique binary string, i.e.,\begin{equation}
    \exists s \in \{0,1\}^{n_A} \st P(S^A=s)=1
\end{equation} we recover sample-splitting. Alternatively, if there are $V$ binary strings among which the probability mass is homogeneously distributed, i.e., \begin{align}
    \exists s^{1},\dots,s^{V} \in \{0,1\}^{n_A} \st P(S^A=s^{j})=1/V \for j=1,\dots,V
\end{align} where $\sum_{j=1}^V s_i^{j} = 1 \textand \sum_{i=1}^{n_A} s_i^{j} = n_A/V$, we recover $V$-fold cross-validation. The sub-empirical distributions $\mathbb{P}_{S,1}^{A}$ and $\mathbb{P}_{S,0}^{A}$ are defined by restricting the empirical distribution $\Pnep$ to the corresponding set,\begin{equation}
    \mathbb{P}_{S^A,j}^A \defeq \frac{1}{n_A(S,j)} \sum_{i=1}^{n_A} \delta_{(X_{A,i},Y_{A,i})} \cdot I(S_i^A=j)
\end{equation} where $n_A(S,j) = \sum_{i=1}^{n_A} I(S_i^A=j)$. Let $S=(S^{A}, S^{\obs})$ be the pair of indicators for each sample, and $(\bXepSj,\bYepSj) \in \R^{\nepj\times(p+1)}$ for $j \in {0,1}$ denote the design matrix and target vector corresponding to the previous sub-empirical distributions, then the in-sample risk on the test set is \begin{equation}
    \Rephatcv(\beta)=\Rephat(\beta,\PSepone)=\norm{\bYepSone - \bXepSone\beta} / \nepj
\end{equation} and the estimator fitted in the in-sample training set is given by equation \eqref{eq:causalRegularizationViaOptHat}, where $\bGlambda$ and $\bZlambda$ are computed on the sub-samples $\{(\bXepSzero,\bXepSzero)\}_{\insample}$
\begin{equation}
    \betahatlambdacv=\betahatlambda(\PSezero,\PSozero)
\end{equation} We define the population and sample selectors as
\begin{alignat}{2}
\tilde{\lambda} &\defeq \argmin_{\lambda\in\Lambda} \tilde{\theta}(\lambda) \quad\st\quad &\tilde{\theta}(\lambda) &\defeq \Rdiff(\beta_\lambda)\qquad\text{Optimal selector}\label{eq:populationselection}\\
\hat{\lambda} &\defeq \argmin_{\lambda\in\Lambda} \hat{\theta}(\lambda) \quad\st\quad &\hat{\theta}(\lambda) &\defeq \E_S|\Rdiffhatcv(\betahatlambdacv)|\qquad\text{Sample selector}\label{eq:sampleselector}
\end{alignat} where the expectation is with respect to the sub-sampling distribution $S$ and where $\Rdiffhatcv(\beta)\defeq\Rehatcv(\beta)-\Rohatcv(\beta)$ is in-sample risk difference on the test set. It holds that the sample and optimal selectors are asymptotically equivalent insofar as $S$ makes both the training and test datasets increase as $n$ increases. This is the case for sample splitting, leave-one-out cross-validation, and V-fold cross-validation insofar as the number of folds grows towards infinity as the sample sizes increase to infinity.

\begin{proposition} If $\Lambda$ is a bounded finite collection, and the distribution of $S$ is such that the training and test datasets increase as the sample size increases \begin{equation}
\min(n(S,1),n(S,0)) \to \infty \as n \to \infty \comma
\end{equation} where $n(S,j)=\min(n_A(S,j),n_0(S,j))$ for $j\in{0,1}$. Then it holds that the sample selection and the optimal selection are asymptotically equivalent in expectation: \begin{equation}
E|\tilde{\theta}(\hat{\lambda})-\tilde{\theta}(\tilde{\lambda})|\to 0 \as n\to\infty\period
\end{equation}\end{proposition}

\begin{proof}
Let \begin{equation}
\ell(\beta) = |R_\Delta(\beta)| \textand \hat{\ell}_{S,1}(\beta) = |R_\Delta^{S,1}(\beta)|.
\end{equation} It follows that
\begin{equation}
|\ell(\beta_\lambda)-E_S \hat{\ell}_{S,1}(\hat{\beta}^{S,0}_\lambda)|\leq E_S\left(|\ell(\beta_\lambda)- \hat{\ell}(\hat{\beta}^{S,0}_\lambda)| + |\ell(\hat{\beta}^{S,0}_\lambda)-\hat{\ell}_{S,1}(\hat{\beta}^{S,0}_\lambda)|\right).
\end{equation} For the first term, we have that \begin{equation}
|\ell(\beta_\lambda)-E_S \hat{\ell}(\hat{\beta}^{S,0}_\lambda)| \leq M \cdot \max\left(\ \norm{\beta_\lambda-\hat{\beta}^{S,0}_\lambda}_1^2\ ,\ \norm{\beta_\lambda-\hat{\beta}^{S,0}_\lambda}_2 \right)
\end{equation} where $M = \max\left(\norm{G_\Delta}_\infty\ ,\ \norm{Z}_2\right)$. For the second term, it holds that
\begin{equation}
|\ell(\hat{\beta}^{S,0}_\lambda)-\hat{\ell}_{S,1}(\hat{\beta}^{S,0}_\lambda)| \leq M_{S,1} \cdot \left(\norm{\beta_{\lambda}-\hat{\beta}^{S,0}_\lambda}_1+\norm{\beta_{\lambda}}_1+1\right)^2
\end{equation} where \begin{equation}
M_{S,1} = \max\left(\ \norm{G_\Delta-\hat{G}^{S,1}_\Delta}_\infty\ ,\ \norm{Z_\Delta-\hat{Z}^{S,1}_\Delta}_\infty\ ,\ \norm{W_\Delta - \hat{W}^{S,1}_\Delta}_\infty\ \right)\period
\end{equation} Thus, in summary, there exists constant $C$ such that \begin{equation}
|\ell(\beta_\lambda)-E_S \hat{\ell}(\hat{\beta}^{S,0}_\lambda)| \leq C \cdot E_S\left[\tilde{M}_{S,1} \cdot \max\left(\ \norm{\beta_\lambda-\hat{\beta}^{S,0}_\lambda}_1^2\ ,\ \norm{\beta_\lambda-\hat{\beta}^{S,0}_\lambda}_1,1\right)\right],
\end{equation} where $\tilde{M}_{S,1} = \max(M_A^{S,1},M_0^{S,1})$ and \begin{equation}
M_A^{S,1}= \max\left(\ \norm{G_A-\hat{G}^{S,1}_A}_\infty\ ,\ \norm{Z_A-\hat{Z}^{S,1}_A}_\infty\ ,\ \norm{W_A - \hat{W}^{S,1}_A}_\infty\ \right).
\end{equation} Note that due to independence between train and test samples, we have that \begin{equation}
E|\ell(\beta_\lambda)-E_S \hat{\ell}(\hat{\beta}^{S,0}_\lambda)| \leq C \cdot E_S\left[ A \cdot B \right],
\end{equation} where \begin{equation}
A = E_{-S}\tilde{M}_{S,1} \textand B= E_{-S}\max\left(\ \norm{\beta_\lambda-\hat{\beta}^{S,0}_\lambda}_1^2\ ,\ \norm{\beta_\lambda-\hat{\beta}^{S,0}_\lambda}_1,1\right),
\end{equation} and $E_{-S}$ denote the expectation over everything except $S$. Since $n(S,1)$ and $n(S,0)$ go to infinity as $n \to \infty$. Then we have that using the exponential tails in and, it holds that \begin{equation}
A\to 0 \textand B\to0 \as n\to\infty.
\end{equation} Consequently, for any $\lambda \in \Lambda$, it holds that \begin{equation}
E|\tilde{\theta}(\lambda)-\hat{\theta}(\lambda)|=E|\ell(\beta_\lambda)-E_S \hat{\ell}(\hat{\beta}^{S,0}_\lambda)| \to 0 \as n\to\infty.
\end{equation}  Furthermore, if $\Lambda$ is finite and bounded, we have that \begin{equation}
\sup_{\lambda \in \Lambda}E|\tilde{\theta}(\lambda)-\hat{\theta}(\lambda)| \to 0 \as n\to\infty.
\end{equation} Finally, since \begin{equation}
0 \leq E|\tilde{\theta}(\hat{\lambda})-\tilde{\theta}(\tilde{\lambda})| \leq 2\sup_{\lambda \in \Lambda}E|\tilde{\theta}(\lambda)-\hat{\theta}(\lambda)|,
\end{equation} we have that \begin{equation}
E|\tilde{\theta}(\hat{\lambda})-\tilde{\theta}(\tilde{\lambda})|\to 0 \as n\to\infty.
\end{equation}
\end{proof}

\section{Additional simulations}\label{sec:additional_simulations}

\subsection{Increasing sample size and fixed dimension}\label{appx:increasing_sample}

\begin{figure}[ht!]
\centering
\includegraphics[width=\linewidth]{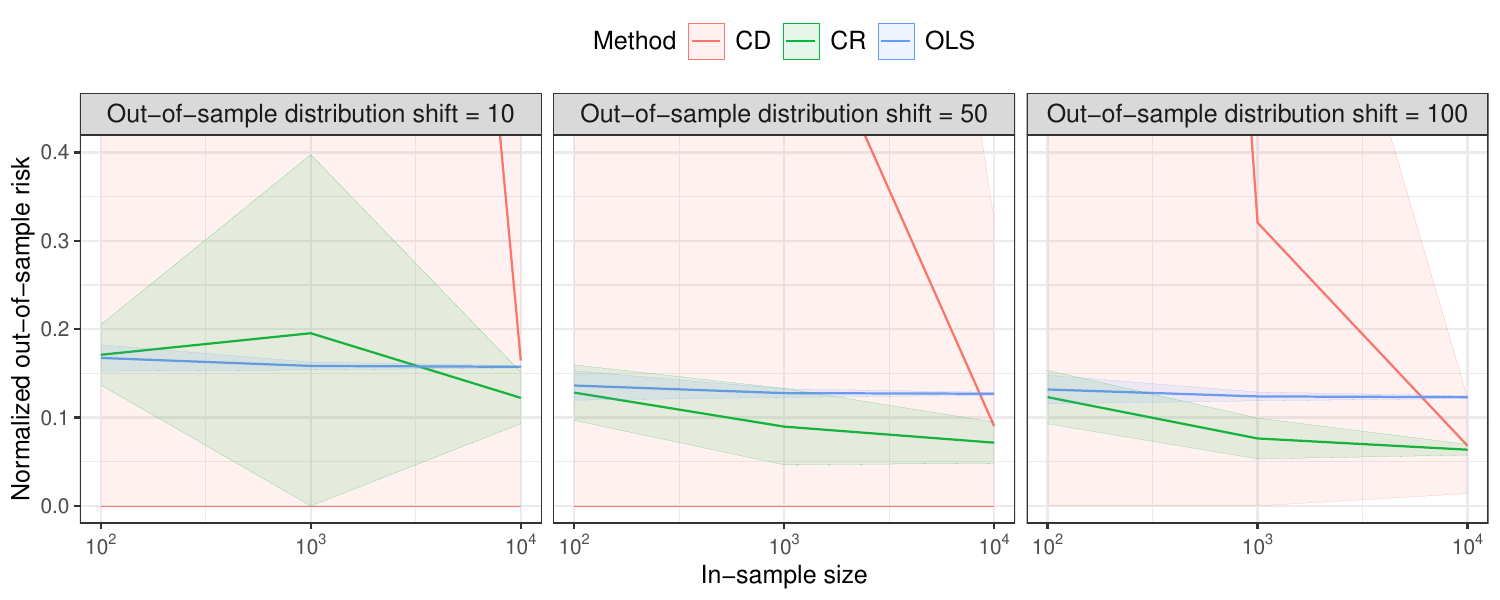}
\caption{Out-of-sample risk for ordinary least-squares, causal dantzig and causal regularization \eqref{eq:causalRegularizationViaOptHat} as the sample size and out-of-sample distribution shift increases. The target variable has been directly shifted. Solid lines indicate average performance, while shaded regions represent one standard deviation across $1000$ trials.}
\label{fig:shift_target_increasing_sample_comparison}
\end{figure}

\begin{figure}[ht!]
\centering
\includegraphics[width=\linewidth]{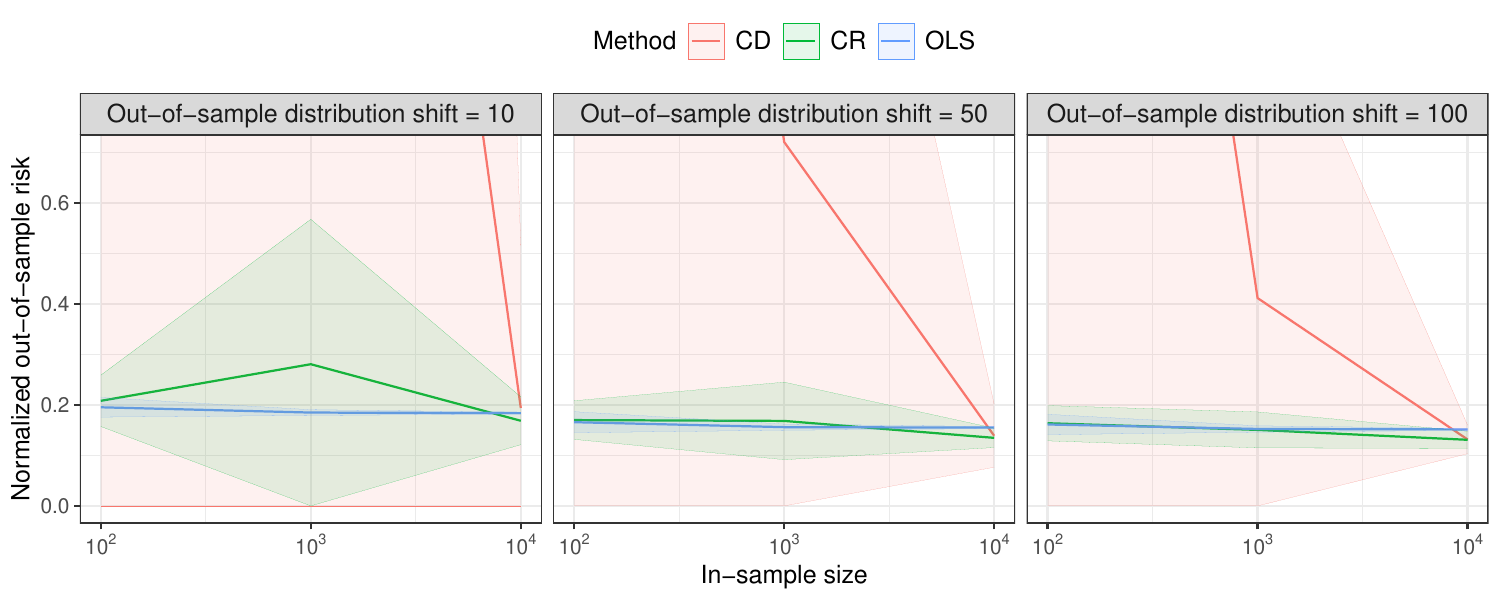}
\caption{Out-of-sample risk for ordinary least-squares, causal dantzig and causal regularization \eqref{eq:causalRegularizationViaOptHat} as the sample size and out-of-sample distribution shift increases. The target variable has been directly shifted. Solid lines indicate average performance, while shaded regions represent one standard deviation across $1000$ trials.}
\label{fig:shift_target2_increasing_sample_comparison}
\end{figure}

We analyse a scenario in which the causal parameters are not identifiable, which impairs the performance of causal Dantzig. To illustrate this, we adopt the same experimental setup described in \zcref{sec:increasing_sample}, but we modify \eqref{eq:dgp} so that the distribution shift directly intervenes on the target variable: \begin{equation}\label{eq:dgp_alt}
A_X \sim \sqrt{\gamma_1}\cdot \mathcal{N}(0_p,I_p), \quad \textand \quad A_Y \sim \sqrt{\gamma_1 \cdot \gamma_2} \cdot \mathcal{N}(0,1),
\end{equation} where $0_p$ denotes a $p$-dimensional column vector of zeroes, and $\gamma_2 \in \{0.25,0.5\}$. \zcref[S]{fig:shift_target_increasing_sample_comparison,fig:shift_target2_increasing_sample_comparison}  display the resulting out-of-sample risk for $\gamma_2=0.25$ and $\gamma_2=0.5$, respectively. As the target variable is shifted more strongly, the in-sample distribution shift is harder to detect. Consequently, the causal Dantzig performs worse, and causal regularization is biased towards OLS. This is evidence by the risk of causal regularization being closer to the risk of OLS in \zcref{fig:shift_target2_increasing_sample_comparison} than in \zcref{fig:shift_target_increasing_sample_comparison}.

\subsection{Increasing dimension}\label{appx:increasing_dimension}

Following the approach in the previous section, we examine a setting where the causal parameters are not identifiable, which reduces the out-of-sample performance of the causal Dantzig. We retain the experimental setup from \zcref{sec:increasing_dimension}, but apply the distribution shift defined in \eqref{eq:dgp_alt}. \zcref[S]{fig:shift_target_increasing_dim_comparison,fig:shift_target2_increasing_dim_comparison} presents the resulting out-of-sample risk for $\gamma_2=0.25$ and $\gamma_2=0.5$ respectively. Analogously to the simulations in \zcref{appx:increasing_sample}, in high-dimensional settings where the target is strongly shifted, i.e. $\gamma_2=0.5$, the causal Dantzig performs badly, evidence by its high variance. Thus, causal regularization chooses parameters closer to OLS. 

% However, in low-dimensional settings where the target variable has not been shifted too strongly, i.e., $\gamma_2=0.25$, causal regularization can provide some minor improvements over OLS.

\begin{figure}[ht!]
\centering
\includegraphics[width=\linewidth]{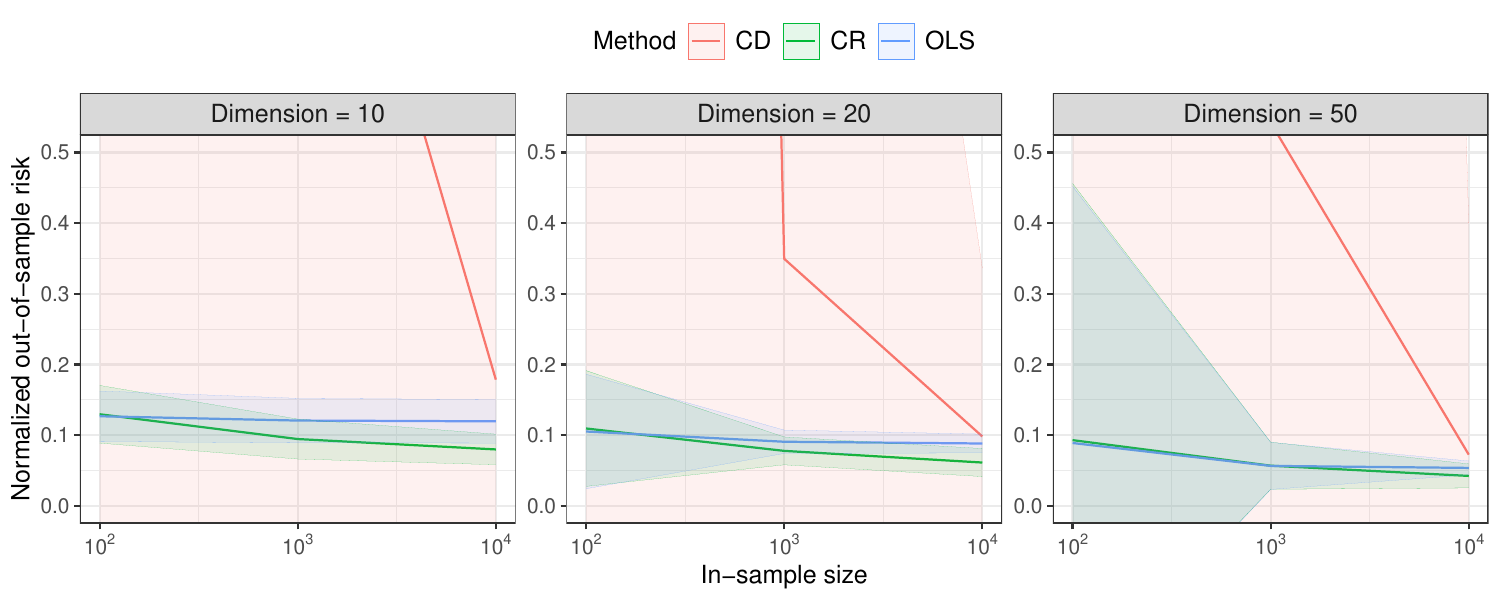}
\caption{Out-of-sample risk for ordinary least-squares, causal dantzig and causal regularization \eqref{eq:causalRegularizationViaOptHat} as the dimension increases. The target variable has been directly shifted. Solid lines indicate average performance, while shaded regions represent one standard deviation across trials. In the third panel, the shaded regions for OLS and causal regularization overlap.}
\label{fig:shift_target_increasing_dim_comparison}
\end{figure}

\begin{figure}[ht!]
\centering
\includegraphics[width=\linewidth]{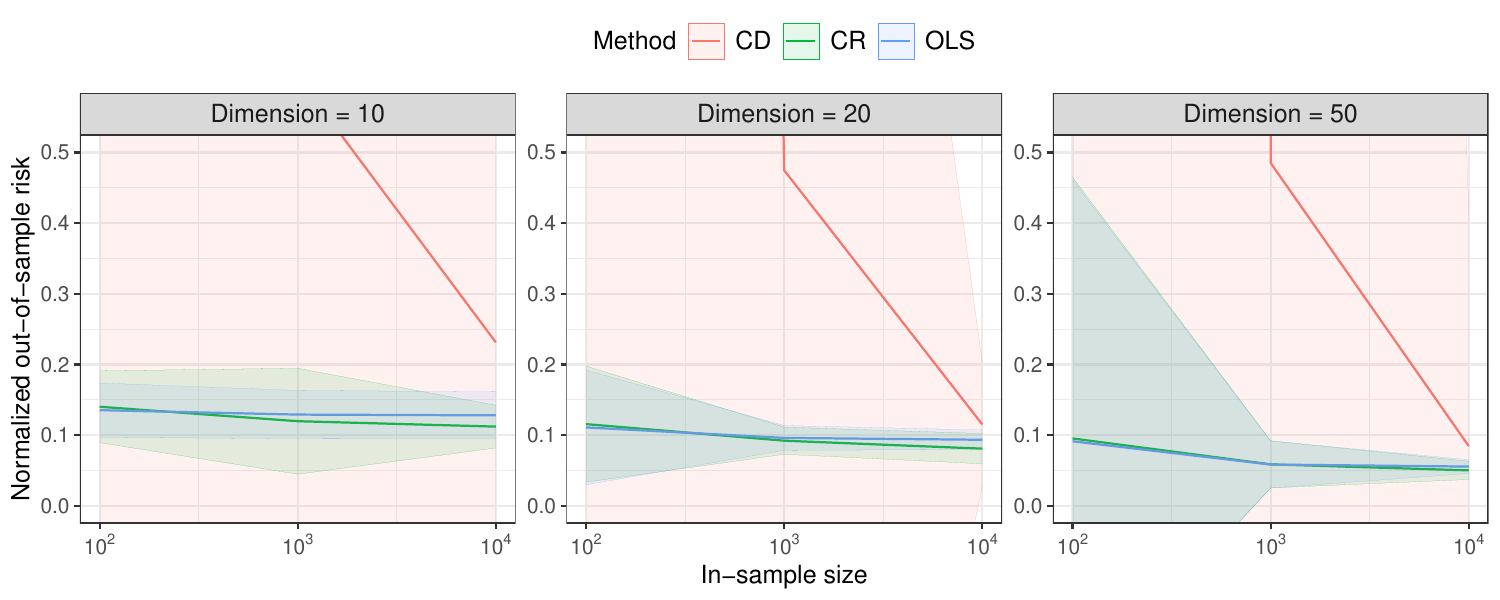}
\caption{Out-of-sample risk for ordinary least-squares, causal dantzig and causal regularization \eqref{eq:causalRegularizationViaOptHat} as the dimension increases. The target variable has been directly shifted. Solid lines indicate average performance, while shaded regions represent one standard deviation across trials. In the third panel, the shaded regions for OLS and causal regularization overlap.}
\label{fig:shift_target2_increasing_dim_comparison}
\end{figure}

\section{Additional experiments}\label{sec:additional_experiments}

In \zcref{sec:fish} and \zcref{sec:genes}, we present two experiments demonstrating that, in the absence of detectable in-sample distribution shift, causal regularization tuned using 10-fold cross-validation selects the OLS estimators with the best out-of-sample performance.

\subsection{Fulton fish market}\label{sec:fish}

We apply our methodology to predict the demand in a fish market from the price \citep{Graddy95,imbensiv}. In equilibrium, it has been hypothesised that \begin{equation}
   \log(\text{quantity}) = \beta_0 + \beta_1 \log(\text{price}) + \epsilon
\end{equation} where quantity is the daily total quantity of fish in pounds, and price is the average daily price in cents.

\begin{figure}[ht]
   \centering
   \includegraphics[width=\linewidth]{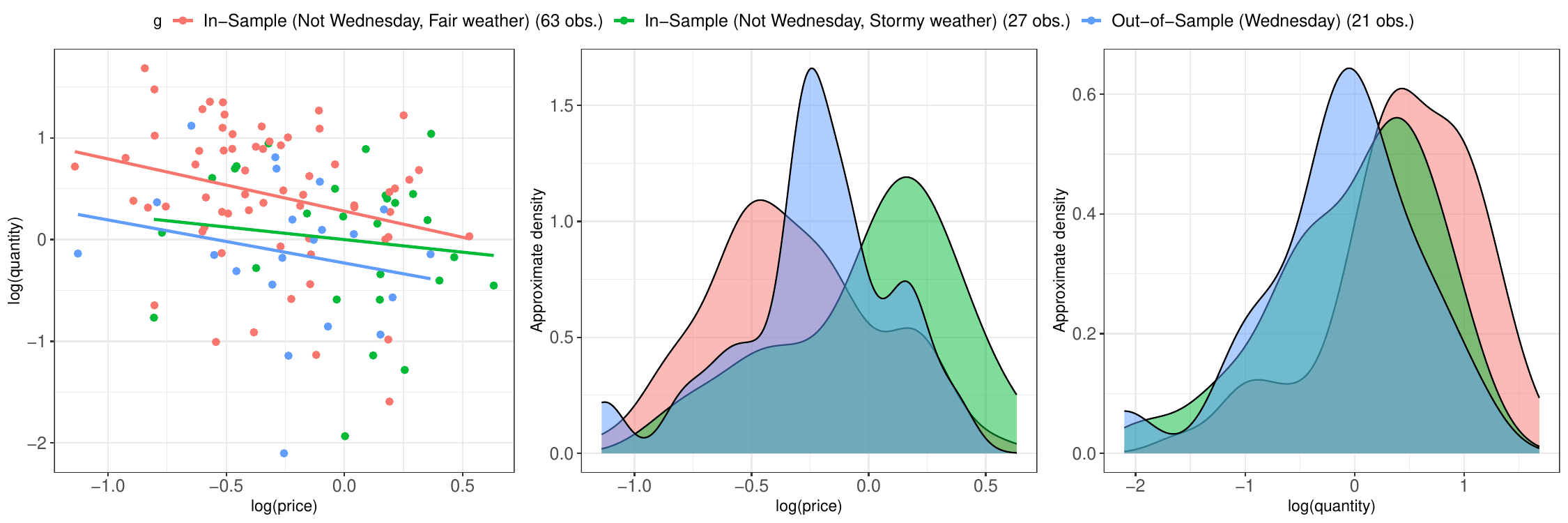}
   \caption{Estimated densities for the amount of fish and its price for each dataset in the Fulton fish market example. This shows that for the covariate, log(price), the observational, fair weather, distribution is indeed different from the shifted, stormy weather, distribution.}
   \label{fig:fish_datasets}
\end{figure}

We are interested in measuring the prediction quality across the week (Mondays, Tuesdays, Wednesdays, and Thursdays). Hence, we split the dataset into a training dataset consisting of Mondays, Tuesdays, and Thursdays, and a test dataset composed of Wednesdays. We further divide the training dataset into two datasets: one for stormy days and one for fair days. Stormy days are those when the wind speed is greater than 18 knots and the wave height is higher than 4.5 feet. \zcref[S]{fig:fish_datasets} visualises the covariate and response for each one of the datasets. 

\zcref[S]{fig:fish_result} presents the in-sample and out-of-sample risks along the causal regularization path across $1000$ resamples. OLS achieves the lowest risk, and the causal regularization method selected through 10-fold cross-validation tends to favour the OLS solution. The substantial variance across resamples arises from the small sample sizes in each dataset. These small samples make the linear system of equations $\tilde{G}_\lambda\beta = \tilde{Z}_\lambda$, which determines the chosen causal regularizer in \eqref{eq:normalization}, ill-posed.

\begin{figure}[ht]
   \centering
   \includegraphics[width=\linewidth]{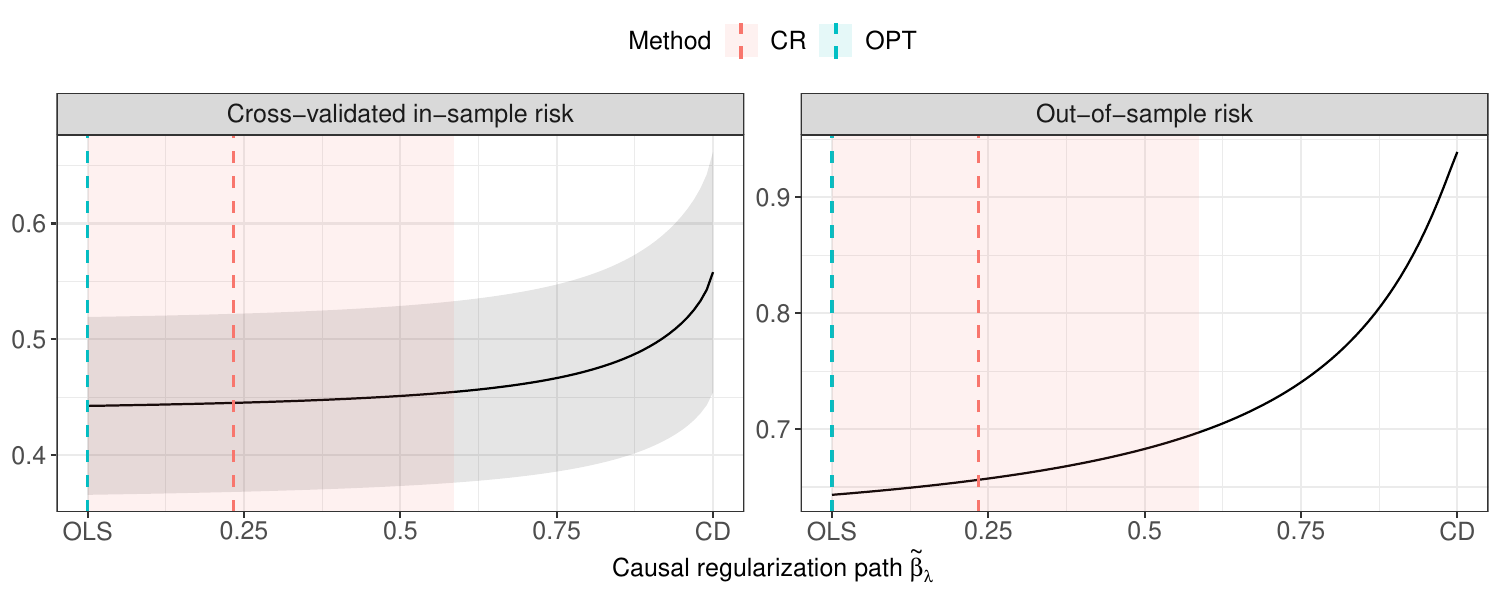}
   \caption{Results for the Fulton fish market in \zcref{sec:fish}. In-sample cross-validation absolute risk difference and out-of-sample risk for the causal regularization path \eqref{eq:causalRegularizationViaOptHat}. The red dashed line indicates the average selected causal regularizer using 10-fold cross-validation. The blue dashed line indicates the optimal choice based on the test dataset. All shaded regions represent one standard deviation across $1000$ resamples. The causal regularization path has been normalized, see \eqref{eq:normalization}.}
   \label{fig:fish_result}
\end{figure}

\subsection{Gene knockout experiments}\label{sec:genes}

We consider a dataset of gene expression in yeast under deletion of single genes \citep{genedata, Meinshausengenes}. There are 262 non-interventional observations, which consist of no gene deletions, and 1479 observations where in each one a different gene is perturbed. In all cases, 6170 genes are measures. The goal is to predict the gene expression of one of the genes based on the others. We take as the target the gene that was not directly intervened on and whose mean was most shifted between the observation and interventional datasets. Analogously, we choose 10 intervened genes whose mean was most shifted between the interventional and observation datasets as the predictors. Finally, the 262 non-interventional observations and half of the interventional observations were used as training data, while the remaining interventional observations were used as testing data. 

\zcref[S]{fig:genes_result} shows the in-sample and out-of-sample risk for the causal regularization path across $1000$ resamples. As in \zcref{sec:fish}, OLS achieves lower risk both in-sample and out-of-sample. In this case, causal regularization tuned via 10-fold cross-validation effectively selects the OLS solution because the variance is minimal. Unlike the experiment in the previous section, the sample size here is large enough to ensure that the linear system of equations $\tilde{G}_\lambda\beta = \tilde{Z}_\lambda$ is well-posed.

\begin{figure}[H]
\centering
\includegraphics[width=\linewidth]{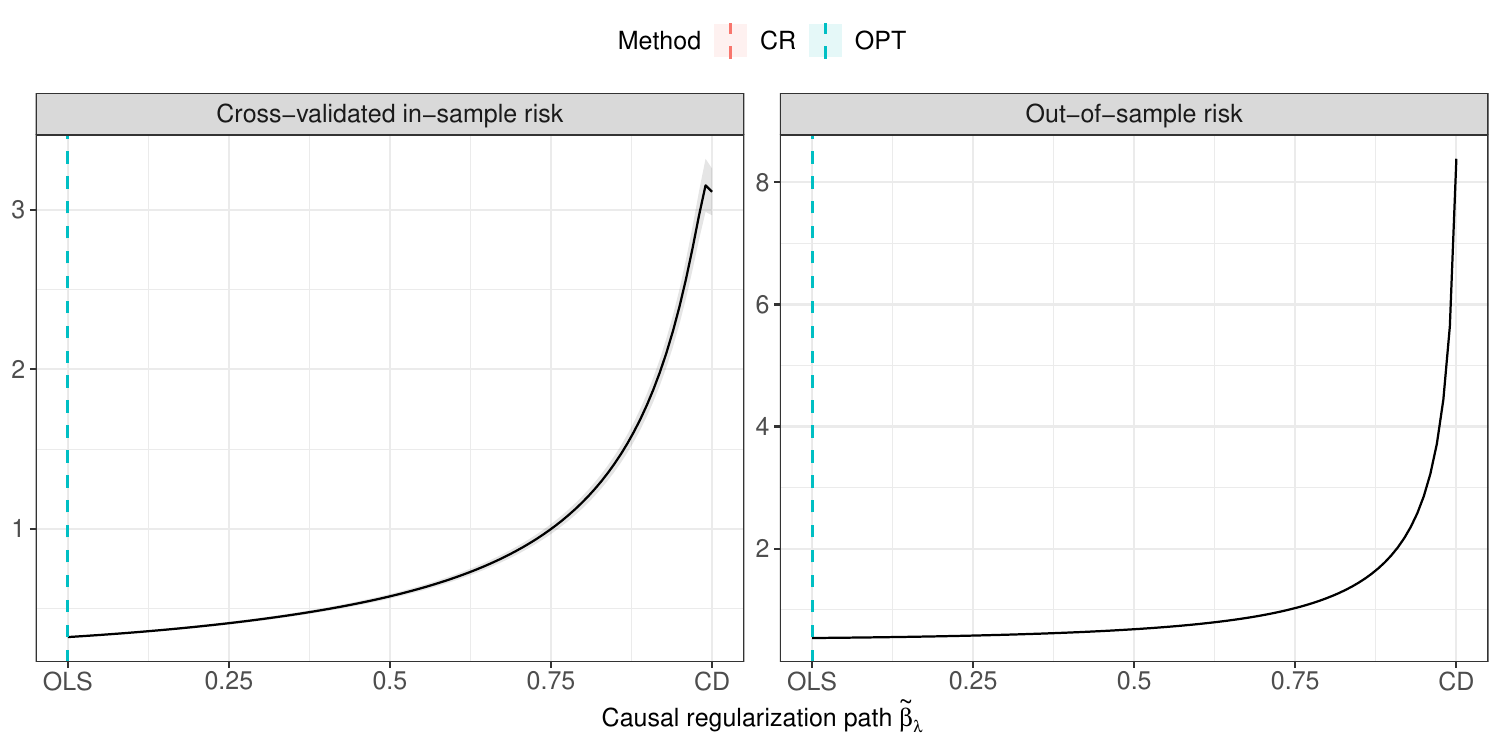}
\caption{Results for the gene knockout experiment in \zcref{sec:genes}. In-sample cross-validation absolute risk difference and out-of-sample risk for the causal regularization path \eqref{eq:causalRegularizationViaOptHat}. The red dashed line indicates the average selected causal regularizer using 10-fold cross-validation. The blue dashed line indicates the optimal choice based on the test dataset. Note that both dashed lines overlap. All shaded regions represent one standard deviation across $1000$ resamples. The causal regularization path has been normalized, see \eqref{eq:normalization}.}
\label{fig:genes_result}
\end{figure}

\etocdepthtag.toc{mtreferences}
\addcontentsline{toc}{section}{References}
\bibliography{biblio.bib}

\end{document}